\documentclass[lettersize,journal]{IEEEtran}
\usepackage{amsmath,amsfonts}
\usepackage{algorithm}    
\usepackage{algpseudocode}
\usepackage{algorithm}
\usepackage{array}
\usepackage[caption=false,font=normalsize,labelfont=sf,textfont=sf]{subfig}
\usepackage{textcomp}
\usepackage{stfloats}
\usepackage{url}
\usepackage{verbatim}
\usepackage{graphicx}
\usepackage{cite}
\usepackage{acronym}
\usepackage[utf8]{inputenc}
\usepackage[T1]{fontenc}
\usepackage{amsmath} 
\usepackage{optidef} 
\usepackage{amsfonts}
\usepackage{amssymb}
\usepackage{bbold}
\usepackage{mathtools}
\usepackage{balance}

\usepackage{hyperref}
\usepackage{xcolor}
\usepackage{empheq}

\usepackage{setspace}	

\begin{document}

\title{Orientation-Aware Control and Trajectory Design for Aerial RIS-Assisted Wireless Communications}

\author{%

\IEEEauthorblockN{
    Abdoul Karim A. H. Saliah$^{1}$, Hajar El Hammouti$^{1}$, Daniel Bonilla Licea$^{1}$, and Giuseppe Silano$^{2}$
}

    \thanks{Copyright (c) 20XX IEEE. Personal use of this material is permitted. However, permission to use this material for any other purposes must be obtained from the IEEE by sending a request to pubs-permissions@ieee.org.}
    
    \thanks{$^1$Abdoul Karim A. H. Saliah, Hajar El Hammouti, and Daniel Bonilla Licea are with the College of Computing, Mohammed VI Polytechnic University, Ben Guerir, Morocco (e-mails: {\tt\small \{abdoul.saliah, hajar.elhammouti, daniel.bonilla\}@um6p.ma}).}
    \thanks{$^2$Giuseppe~Silano is with the Department of Power Generation Technologies and Materials, Ricerca sul Sistema Energetico (RSE) S.p.A., Milan, Italy, and also with the Department of Cybernetics, Czech Technical University in Prague, Prague, Czech Republic (email: {\tt\small giuseppe.silano@fel.cvut.cz}).}
    \thanks{This work was partially funded by the European Union grant no. DCI-PANAF/2020/420-028, through the African Research Initiative for Scientific Excellence (ARISE), pilot programme, by the research fund for the Italian Electrical System (Ricerca di Sistema) through the decree n. 388 of November 6th, 2024, by the GAČR project no. 26-22606S, and the CTU grant no.~SGS26/077/OHK3/1T/13. ARISE is implemented by the African Academy of Sciences with support from the European Commission and the African Union Commission. The contents of this document are the sole responsibility of the author(s) and can under no circumstances be regarded as reflecting the position of the European Union, the African Academy of Sciences, and the African Union Commission.}%
}

\markboth{IEEE TRANSACTIONS ON VEHICULAR TECHNOLOGY. PREPRINT VERSION. ACCEPTED SEPTEMBER, 2026}%
{}



\maketitle

\begin{abstract}
    Aerial Reconfigurable Intelligent Surfaces (ARISs) can enhance wireless connectivity in obstructed environments by combining reconfigurable metasurfaces with aerial mobility. However, in practical deployments, the Reconfigurable Intelligent Surface (RIS) is rigidly mounted on a multi-rotor Unmanned Aerial Vehicle (UAV), which creates coupled position–orientation dynamics that directly affect communication quality. Since RIS performance is highly sensitive to orientation, point-mass or orientation-invariant models may lead to optimistic performance estimates. 
    This paper proposes an orientation-aware ARIS framework that explicitly accounts for UAV dynamics, actuation limits, and orientation-dependent channels. The proposed approach combines a communications-guided reference trajectory planner with a Nonlinear Model Predictive Control for dynamically feasible trajectory tracking and favorable RIS orientation, while employing lightweight beamforming and RIS phase-shift updates. Simulations show that accounting for orientation improves user throughput by $13.77\%$, while the proposed joint communication updates improve network throughput by $28\%$ over phase-shift-only optimization and by at least a factor of $2.5$ over beamforming-only optimization. These results highlight the importance of orientation-aware modeling and joint communication-control design in ARIS-assisted wireless systems.
\end{abstract}






\section{Introduction}
\label{sec:introduction}
 
\IEEEPARstart{U}{nmanned} Aerial Vehicles (UAVs) are widely recognized as a key enabling technology for future 6G integrated terrestrial and non-terrestrial networks, owing to their high mobility, flexible deployment, and ability to establish dominant line-of-sight (LoS) communication links~\cite{Wang2023AntiJamming,Ray2022SAGIN}. These characteristics make UAVs particularly attractive for on-demand coverage extension, capacity enhancement, and mission-critical services in scenarios such as disaster response, dense urban environments, and network offloading~\cite{Dai2023UAV,Bonilla2024RoboticsAndComm}. As a result, UAVs have been extensively studied as aerial base stations, relays, and mobile edge computing platforms~\cite{Vaezi20225G6G}.

Despite these advantages, UAV-assisted communications face significant challenges in obstructed and dense environments, where severe path loss and multipath fading can degrade link quality~\cite{Eskandari2023LoS,Savkin2023MultiUAV}. Addressing these issues has recently motivated strong interest in Reconfigurable Intelligent Surfaces (RIS), a promising technology for shaping and controlling wireless propagation environments~\cite{Liu2021RIS, Yang2024RIS5G}. A RIS consists of a planar array of programmable meta-elements whose electromagnetic response can be dynamically adjusted, typically by controlling phase shifts, to manipulate incident waves through reflection, transmission, or refraction. A prominent subclass is the Intelligent Reflecting Surface (IRS), which specializes in passive beamforming \cite{Wu2024IS_IRS_RIS}. By appropriately configuring these elements, RISs can enhance signal coverage and reliability, especially in non-LoS conditions~\cite{Cao2021RIS,Peng2024BeamformingRIS}. For the sake of simplicity, we use the broader term RIS throughout the paper to refer specifically to passive IRS designs.

An Aerial RIS (ARIS) is a RIS mounted on a UAV that combines intelligent signal manipulation with aerial mobility. This integration enables the RIS to be repositioned in three-dimensional space, providing additional degrees of freedom for improving communication performance. Several recent studies have explored ARIS-enabled networks for applications such as age-of-information minimization in IoT systems, energy-efficient mobile edge computing, and physical-layer security enhancement~\cite{Sherman2023AoI,Arzykulov2024Aerial}. These works typically focus on jointly optimizing the UAV trajectory, RIS phase shifts, and, in some cases, transmitter beamforming or resource allocation.

However, a significant realism gap remains in much of the existing ARIS literature. Most contributions model the UAV as a freely moving point mass, implicitly assuming that position and orientation can be controlled independently. In this work, we focus on ARIS implementations based on multi-rotor UAVs, which are the most commonly adopted platforms in practice due to their hovering capability and precise maneuverability~\cite{ribeiro2025Energy}. For such vehicles, translation and orientation are inherently coupled~\cite{Leutenegger2016Flying}: horizontal motion necessarily requires vehicle tilting, which directly alters the orientation of a rigidly mounted RIS. Since RIS performance is highly sensitive to its pose~\cite{Lu2021AerialRIS}, neglecting this coupling can lead to inaccurate performance predictions and overly optimistic conclusions. Moreover, many studies further simplify the problem by assuming isotropic RIS behavior~\cite{Lu2021AerialRIS}, an assumption that does not reflect the directional nature of practical RIS implementations.


Motivated by these observations, this work addresses the problem of designing ARIS trajectories and communication parameters while explicitly accounting for UAV dynamics, actuation limits, and the orientation-dependent behavior of RIS-assisted channels.

To this end, we consider an ARIS-assisted terrestrial network in which a multi-rotor UAV–mounted RIS evolves from a specified initial state to a target final state while supporting ground users. The objective is to maximize network throughput subject to communication Quality-of-Service (QoS) requirements, UAV dynamic constraints, and actuation limits. The resulting problem tightly couples mobility, control, and communication variables, leading to a challenging nonconvex optimization problem. To manage this complexity, we adopt a structured solution approach that separates long-horizon trajectory generation from constraint-aware tracking, allowing communication performance and vehicle feasibility to be jointly addressed within a unified framework.



\subsection{Related work}

Research on ARIS has grown rapidly in recent years, with most studies focusing on communication-centric optimization problems involving UAV trajectory design, transmitter beamforming, and RIS phase-shift configuration~\cite{wang2023irs , wu2020irs}. These works consistently demonstrate that aerial mobility can significantly enhance coverage, adaptability, and spectral efficiency compared to static RIS deployments~\cite{alsenwi2025risuav}. However, many existing approaches rely on simplifying assumptions that limit their applicability to realistic ARIS platforms. 

The main limitations of the current literature can be broadly grouped into three areas: (i) simplified modeling of UAV motion and actuation, (ii) incomplete characterization of orientation-dependent RIS–channel interactions, and (iii) solution strategies that rely on strong approximations or learning-based heuristics with limited interpretability and deployment guarantees.

\textbf{Simplified UAV motion modeling:} A large portion of the ARIS literature models the UAV as a freely positionable point mass with either fixed or implicitly controllable orientation~\cite{Mei2021Joint, Mei2022Trajectory}. This abstraction neglects the fact that, for UAV platforms commonly considered in ARIS applications, translational motion is coupled with attitude dynamics. In practice, horizontal motion requires vehicle tilting, implying that changes in position are generally accompanied by changes in orientation, particularly during accelerations or trajectory maneuvers. Ignoring this coupling can lead to trajectories and orientation profiles that are difficult or impossible to realize on a physical platform, potentially degrading the communication performance when implemented on real systems. While such simplified models are useful for high-level communication analysis, they do not fully capture the constraints imposed by UAV dynamics and actuation limits. For instance, in~\cite{Mei2021Joint}, an ARIS is used as an edge computing server with energy minimization via joint trajectory, task offloading, and cache optimization, while~\cite{Mei2022Trajectory} jointly optimizes trajectory and phase shifts to minimize energy consumption and maximize data rate. Both works model the UAV as a freely positionable point mass, ignoring dynamic constraints.

\textbf{Incomplete channel and RIS modeling:} Most ARIS studies adopt channel models that depend only on the positions of the transmitter, receiver, and RIS, while assuming a fixed or orientation-independent surface response~\cite{Duo2023Joint}. This assumption is consistent with static RIS deployments but becomes questionable when the RIS is mounted on a moving aerial platform. Even when attitude variations are moderate, changes in orientation can significantly affect the effective reflection pattern and link quality. Additionally, many works assume an isotropic RIS response~\cite{Lu2021AerialRIS,Li2024JointTrajBeam}, implicitly treating the surface as reflecting energy uniformly in all directions. Practical RIS implementations, however, exhibit directional and anisotropic radiation characteristics that depend on both position and orientation. More recent models that explicitly account for this anisotropy~\cite{Liu2024ARIS} provide a more accurate representation of ARIS-assisted communication and motivate the need for orientation-aware formulations. For example, in~\cite{Duo2023Joint}, an ARIS forwards devices’ data to a backhaul UAV with the objective of maximizing energy efficiency, but assumes fixed UAV orientation and isotropic RIS elements. In~\cite{Liu2024ARIS}, the RIS is modeled as anisotropic; however, the ARIS is restricted to hovering, which limits its potential.

\textbf{Solution strategies and computational considerations:} The joint optimization of UAV trajectory, RIS phase shifts, and beamforming is typically addressed using either convex-approximation techniques or learning-based methods. Approaches based on Successive Convex Approximation~\cite{Adam2024SecureUAVRIS} reduce computational complexity by linearizing nonconvex terms and decoupling variables, but they generally provide locally optimal solutions whose quality depends on initialization and modeling assumptions~\cite{wu2019irs}. More recently, Deep Reinforcement Learning (DRL) has been applied to ARIS optimization problems to handle high-dimensional and nonlinear dynamics~\cite{Guo2023DrlNomaRIS,Tang2023SecureRIS}. While DRL offers flexibility in exploring complex solution spaces, its deployment is often hindered by substantial offline training requirements and limited interpretability. Moreover, stability and constraint satisfaction are typically not guaranteed by design, which is a critical aspect for aerial platforms operating under actuation limits \cite{brunke2022safe,cheng2019safe}.

Among the works most closely related to this paper are those by Eskandari et al.~\cite{Eskandari2023MpcARIS} and Li et al.~\cite{Li2025ARIS}, which attempt to incorporate more realistic vehicle behavior. Eskandari et al. employ a non-holonomic kinematic model combined with MPC-based planning; however, the model is tailored to fixed-wing platforms~\cite{Lee2021platooning}  and does not capture the coupled translational–rotational dynamics relevant to multi-rotor ARIS configurations or their impact on the communication channel. Li et al. include rotational motion and orientation-aware radiation patterns within a DRL framework, representing an important step toward realism. Nevertheless, their formulation does not explicitly model actuator constraints or second-order rotational dynamics, and the learning-based approach inherits the typical challenges associated with DRL-based solutions.

In summary, existing ARIS studies often rely on simplified UAV motion models, neglect orientation-dependent RIS behavior, or adopt solution methods that are difficult to deploy within a constraint-aware control architecture. These limitations motivate the need for an approach that jointly considers realistic UAV dynamics, orientation-aware channel modeling, and optimization methods that can be naturally embedded within a control framework for reliable ARIS operation.



\subsection{Contributions}

Motivated by the limitations identified in existing ARIS literature, this paper proposes a communications-aware control framework for UAV-mounted ARIS platforms operating in obstructed environments. The framework explicitly couples vehicle motion, orientation evolution, and wireless communication performance, enabling the joint design of mobility and communication strategies beyond simplified point-mass or orientation-invariant models. In particular, we consider an ARIS acting as an aerial relay and address the joint optimization of UAV trajectory, beamforming, and RIS phase shifts under realistic dynamic and communication constraints. 

The main contributions of this work are summarized as follows:

\begin{enumerate}
    \item We develop an integrated model that combines full UAV translational and rotational dynamics, actuation limits, and an orientation-dependent, non-isotropic ARIS channel characterization based on angle-dependent aperture gain. This model captures physical effects neglected in most existing ARIS formulations.
    \item The coupled optimization of trajectory, beamforming, and RIS phase shifts is decomposed into tractable subproblems, enabling closed-form solutions for beamforming and phase-shift updates while preserving their coupling with UAV motion. This structure significantly reduces computational complexity compared to monolithic or learning-based formulations.
    \item We introduce a two-stage framework consisting of a communications-aware trajectory planner and a NMPC tracker that enforces UAV dynamics, actuation limits, and data rate QoS constraints during execution. The planner addresses the global communication-geometry problem, while the NMPC ensures dynamically feasible motion, favorable ARIS orientation, and reliable link quality.
    \item Extensive simulations demonstrate that explicitly accounting for UAV orientation and dynamics improves user throughput by $13.77\%$, while the proposed joint optimization increases network throughput by $28\%$ over phase-shift-only optimization and by a factor of at least $2.5$  over beamforming-only optimization, all while ensuring accurate trajectory tracking under realistic actuation constraints.
\end{enumerate}

Compared to existing ARIS approaches, the proposed framework advances the state-of-the-art in three key aspects: (i) it moves beyond point-mass and kinematic UAV models by explicitly accounting for coupled translational–rotational dynamics; (ii) it incorporates orientation-dependent, non-isotropic ARIS channel effects that are typically ignored; and (iii) it provides a computationally efficient and control-oriented solution that can be embedded within a feedback architecture, avoiding the training overhead and deployment concerns associated with learning-based methods. To the best of the authors' knowledge, few ARIS studies combine orientation-aware channel modeling, multi-rotor UAV dynamics, explicit actuation limits, and NMPC-based tracking within a single deployable framework. 


\textbf{Notation:} Scalars are denoted by italic lowercase letters (e.g., $y$), vectors by bold lowercase letters (e.g., $\mathbf{y}$), and matrices by bold uppercase letters (e.g., $\mathbf{Y}$). The transpose and Hermitian transpose of a matrix $\mathbf{Y}$ are denoted by $\mathbf{Y}^\top$ and $\mathbf{Y}^\dagger$, respectively. The modulus of a complex number is denoted by $|\cdot|$, and $\|\cdot\|_2$ denotes the Euclidean ($\ell^2$) norm. The imaginary unit is denoted by $j$. The operator $[\cdot]_3$ denotes the third component of a three-dimensional vector. The unit quaternion $\mathbf{q}$ belongs to the 3-sphere $\mathbb{S}^3 = \{ \mathbf{q} \in \mathbb{R}^4 : \|\mathbf{q}\|_2 = 1 \}$, which represents the set of unit-norm quaternions used to parametrize three-dimensional rotations~\cite{Bayro2021Quaternion}. The corresponding rotation matrix $\mathbf{R}(\mathbf{q})$ belongs to the Special Orthogonal Group $\mathrm{SO}(3) = \{ \mathbf{R} \in \mathbb{R}^{3 \times 3} : \mathbf{R}^\top \mathbf{R} = \mathbf{I}, \det(\mathbf{R}) = 1 \}$, which denotes the set of all proper rotation matrices in three-dimensional space. The symbols $\times$, $\angle(\cdot)$, $\left\lfloor \cdot \right\rfloor$, and $\circ$ denote the vector cross product, the phase of a complex number, the floor of a real number, and the quaternion product, respectively.




\section{System and Communication Models}
\label{SystAndComModel}

This section presents the system modeling adopted in the paper. We first describe the ARIS-assisted communication scenario, including the network architecture and coordinate frames (Section \ref{SystemModel}). We then introduce an orientation-aware propagation geometry of the ARIS-assisted terrestrial communications (Section \ref{OrientationARISProp}), followed by the channel model (Section \ref{ChannelModel_Sys}). Finally, we present the UAV dynamic model used for trajectory design and control (Section \ref{UAV_DM}).

\begin{figure}[tb]
    \centering
    \includegraphics[width=1\linewidth]{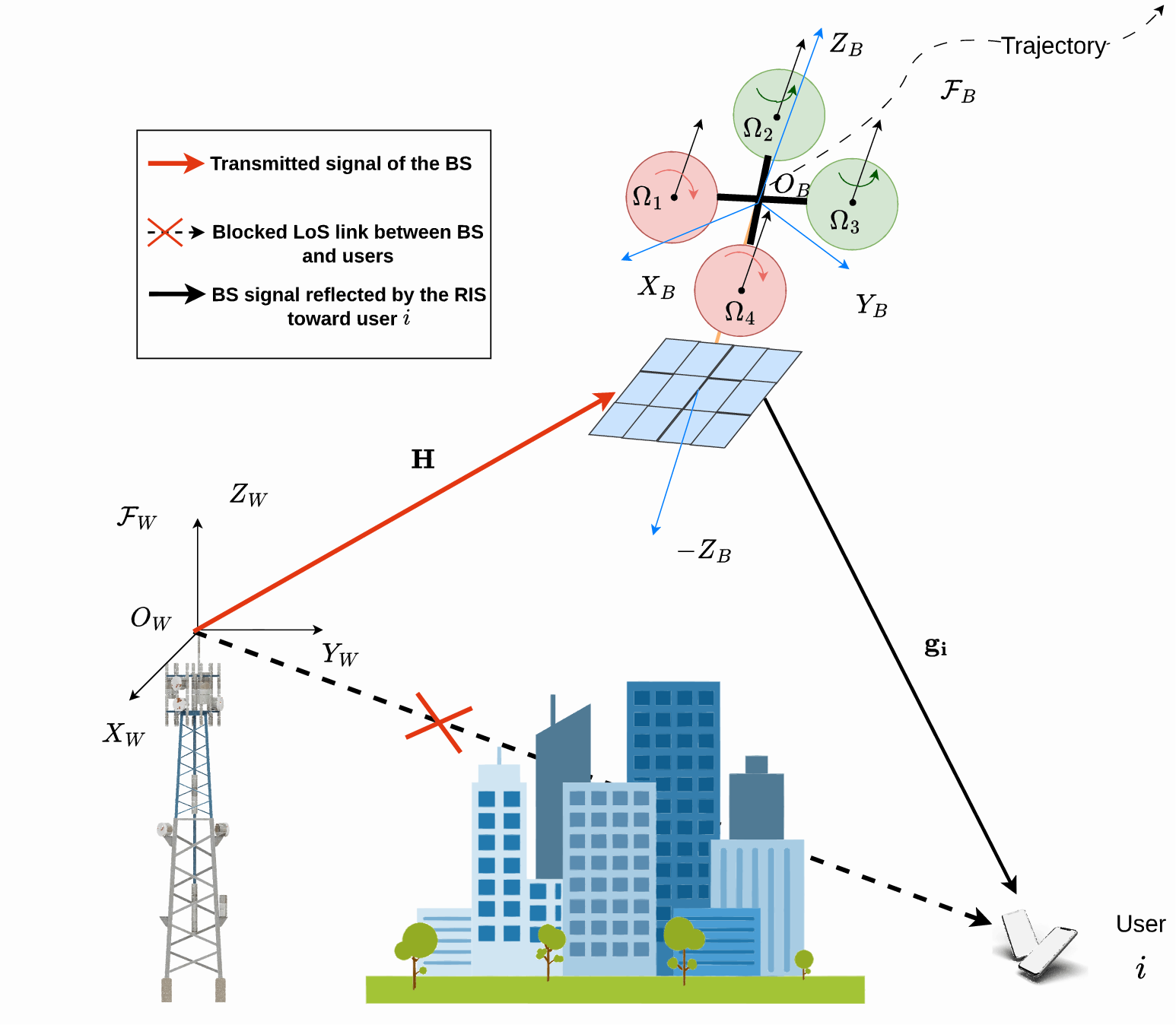}
    \vspace{-2.25em}
    \caption{ARIS-assisted communication system architecture: a RIS rigidly mounted on a multi-rotor UAV relays signals between the base station and ground users. The UAV trajectory is expressed in the world (inertial) reference frame $\mathcal{F}_W$, and a RIS orientation in the UAV body-fixed frame $\mathcal{F}_B$.}
    \label{Sys-Archi}
    \vspace{-0.5cm}
\end{figure}

\begin{table}[tb]
    \centering
    \footnotesize
    \caption{Notation List}
    \vspace{-1em}
    \label{TabNotationSystem}
    \renewcommand{\arraystretch}{1.0} 
    \setlength{\tabcolsep}{4pt}
    \begin{tabular}{l|p{6.25cm}}
    \hline
    \textbf{Notation} & \textbf{Description} \\ \hline
    \multicolumn{2}{c}{\textbf{Communication}} \\ \hline
    $\mathcal{I}, \mathcal{N}, \mathcal{M}$ & User, BS-antenna, and RIS-element index sets  \\
    $I, N, M$ & Number of users, BS antennas, and RIS elements \\
    $\mathbf{H}, \mathbf{g}_i, \boldsymbol{\Theta}$ & BS-RIS and RIS-user $i$ channels, and phase-shift \\
    $\mathbf{p}^{\text{BS}}, \mathbf{p}_i, \mathbf{p}^\text{v}$ & Position of BS, of user $i$, and of a virtual user located at the barycenter of the users' positions \\
    $\mathbf{f}, G^{\mathrm{ap}}_i$ & BS beamforming vector and RIS aperture gain of $i$\\
    $\eta_i, R_i, R_{\min}$ & SNR and data rate of $i$, and minimum QoS data rate  \\
    $P_0, B_0, \sigma^2$ & BS transmit power, bandwidth, and noise power \\ \hline
    \multicolumn{2}{c}{\textbf{UAV Dynamic Model}} \\ \hline
    $\mathbf{p}, \mathbf{q}, \mathbf{v}, \boldsymbol{\omega}$ & Position, orientation, linear and angular velocity \\
    $m, g, \mathbf{J}$ & UAV mass, gravitational constant, and inertia matrix \\
    $u_T, \boldsymbol{\tau}, \mathbf{\Omega}$ & Total thrust, control torque, and rotor speeds \\
    $\underline{\Omega}, \bar{\Omega}, \bar{v}$ & Rotor speed limits and maximum UAV velocity \\
    $\mathbf{x}, \mathbf{u}, \mathbf{h}(\cdot)$ & UAV state, control input, and dynamics \\ \hline
    \multicolumn{2}{c}{\textbf{NMPC}} \\ \hline
    $T_s, N_p$ & Sampling time and prediction horizon \\
    $\mathbf{x}_\ell, \mathbf{u}_\ell, \mathbf{x}_{d,\ell}, s_{i,\ell}$ & State, control, reference state, and QoS slack at $\ell$ \\
    $d_q(\cdot), \mathcal{A}$ & Geodesic distance and flight region \\ \hline
    \end{tabular}
    \vspace{-0.30cm}
\end{table}



\subsection{System and coordinate frames}
\label{SystemModel}

We consider a downlink wireless communication system in which a terrestrial BS serves multiple Ground Users (GUs) located in an obstructed environment. Since the direct BS--GU links are blocked, a UAV-mounted RIS is deployed as an aerial relay to reflect the incident signals toward the users. The RIS is rigidly attached to the UAV such that its geometric center coincides with the vehicle's Center of Mass (CoM). The UAV is required to move from an initial state to a target final state while maintaining reliable communication with the users over the mission duration $T$. An overview of the system architecture is shown in Fig.~\ref{Sys-Archi}, and a notation table is given in Table \ref{TabNotationSystem}.

The BS is equipped with a Uniform Linear Array (ULA) consisting of $N$ antenna elements, indexed by the set $\mathcal{N} = \{1,\dots,N\}$, and serves $I$ ground users indexed by $\mathcal{I} = \{1,\dots,I\}$. The UAV carries a RIS Uniform Planar Array (UPA) composed of $M = M_H M_V$ reflecting elements, indexed by the set $\mathcal{M} = \{1,\dots,M\}$.

\begin{figure}[t]
    \centering
    \includegraphics[width=0.9\linewidth]{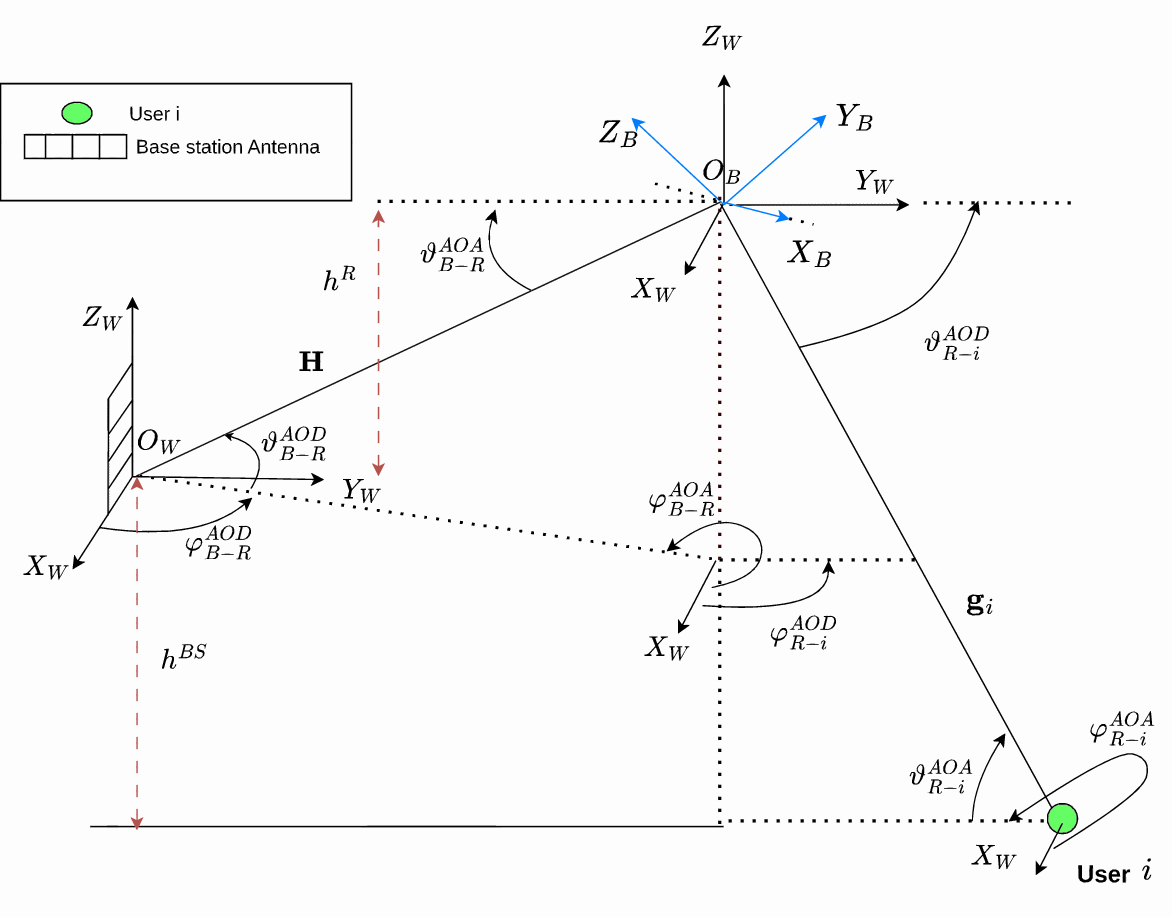}
    \vspace{-1.5em}
    \caption{Geometric model of the ARIS-assisted communication system with orientation-aware coordinate frames and signal directions.}
    \label{fig:Model-Geo}  
    \vspace{-1.7em}
\end{figure}

Two coordinate frames are defined (see Fig.~\ref{fig:Model-Geo}): a world (inertial) frame $\mathcal{F}_W$ and a body-fixed frame $\mathcal{F}_B$. The origin of $\mathcal{F}_W$ is located at the BS reference point, while the origin of $\mathcal{F}_B$ coincides with the UAV CoM. The positions of the BS, the UAV (ARIS), and user $i$ expressed in $\mathcal{F}_W$ are denoted by $\mathbf{p}^{\text{BS}} = [0,\,0,\,0]^\top$, $\mathbf{p} = [p_x,\,p_y,\,p_z]^\top$, $\mathbf{p}_i = [p_{x,i},\,p_{y,i},\,-h^{\mathrm{BS}}]^\top$, respectively, where $h^{\mathrm{BS}}$ denotes the BS height. Since the origin of $\mathcal{F}_W$ is placed at the BS reference point, ground-level users lie at $z=-h^{\text{BS}}$ in this coordinate system.

The UAV orientation is represented by a unit quaternion $\mathbf{q} \in \mathbb{S}^3$, which provides a singularity-free representation of three-dimensional attitude and is well suited for modeling rotational dynamics~\cite{Bayro2021Quaternion}. The corresponding rotation matrix $\mathbf{R}(\mathbf{q}) \in \mathrm{SO}(3)$ maps vectors from the body frame $\mathcal{F}_B$ to the world frame $\mathcal{F}_W$ and will be used in the subsequent channel and dynamics modeling.

To serve multiple users, we adopt a Time Division Multiple Access (TDMA) protocol, where each user is allocated a time slot with the same duration. We assume that the coherence time of the channel is larger than the period of the TDMA frame. During the transmission slot, the scheduled user occupies the entire bandwidth $B_0$, a single beamforming vector is employed, and inter-user interference is absent~\cite {Ghasemi2024RPA,Ghasemi2025RAU}.


\subsection{Orientation-aware ARIS propagation geometry}
\label{OrientationARISProp}

The position $\mathbf{p}_n^\text{BS}$ of an individual antenna element $n$ in the world frame $\mathcal{F}_W$, and the position $\mathbf{p}^\text{R}_m$ of the $m$-th ARIS element in the UPA in the body frame $\mathcal{F}_B$, are defined as in \cite{Saliah2025Harnessing}.

An important component in RIS channel modeling is the wave vector \(\mathbf{s}(\vartheta, \varphi)\), which defines the direction of propagation and phase progression of a plane wave in the world frame~\cite{bjornson2017mimo}. For a generic communication link \(W \to Z\), the wave vector is expressed as
\begin{equation}
    \mathbf{s}(\vartheta_{W-Z},\varphi_{W-Z}) =
    \begin{bmatrix}
    \cos\vartheta_{W-Z}\cos\varphi_{W-Z} \\
    \cos\vartheta_{W-Z}\sin\varphi_{W-Z} \\
    \sin\vartheta_{W-Z}
    \end{bmatrix},
\end{equation}
where $\vartheta_{_{W-Z}}$ and $\varphi_{_{W-Z}}$ are the elevation and azimuth angles characterizing the Angle of Departure (AoD) or Angle of Arrival (AoA) between nodes $W$ and $Z$, respectively.

In our system,  we model the links between the RIS and BS and between the RIS and user using the following notation: \(B\) for the BS, \(R\) for the ARIS, and \(i\) for a ground user. This gives rise to two main links: \(B \to R\) and \(R \to i\).

The path difference\footnote{The path difference is the propagation distance offset between the path involving the $m$-th RIS element and the reference path through the RIS geometric center. For the incident wave, between the BS–element and the BS–center paths; for the reflected wave, between the element–receiver path and the center–receiver paths. It accounts for the relative phase shift introduced by each element's spatial displacement from the RIS center.} is obtained by projecting the global position of the $m$-th element onto the corresponding wave vector. For the BS-to-ARIS link, this additional path length is expressed as
\begin{equation}
    \resizebox{0.910\hsize}{!}{$%
    \Delta\ell_{m}\left(\vartheta^{\text{AoA}}_{B-R},\varphi^{\text{AoA}}_{B-R},\mathbf{q}\right) = \mathbf{s}(\vartheta^{\text{AoA}}_{B-R},\varphi^{\text{AoA}}_{B-R}) \cdot \left( \mathbf{R}(\mathbf{q})  \mathbf{p}^\text{R}_{m} \right),
    $}%
    \label{eq:path_diff_bs_ris}
\end{equation}
where \(\mathbf{R}(\mathbf{q}) \mathbf{p}^{R}_{m}\) is the position of the \(m\)-th RIS element rotated into the world frame $\mathcal{F}_W$, and  $\left(\vartheta_{B-R}^{\text{AoA}}, \varphi_{B-R}^{\text{AoA}}\right)$ are the elevation and the azimuth AoAs of the link $B-R$.

Similarly, for the ARIS-to-user \(i\) link, the path difference from the \(m\)-th element to the user is
\begin{equation}
    \Delta\ell_{m,i}\left(\vartheta^{\text{AoD}}_{R-i},\varphi^{\text{AoD}}_{R-i},\mathbf{q}\right) = \mathbf{s}(\vartheta^{\text{AoD}}_{R-i},\varphi^{\text{AoD}}_{R-i}) \cdot \left( \mathbf{R}(\mathbf{q})  \mathbf{p}^\text{R}_{m} \right),
    \label{eq:path_diff_ris_user}
\end{equation}
where $\left(\vartheta^{\text{AoD}}_{R-i},\varphi^{\text{AoD}}_{R-i} \right)$ are the elevation and azimuth AoDs of the link $R-i$.



\subsection{Channel model}
\label{ChannelModel_Sys}

The channels and the model geometry of the ARIS are illustrated in Fig. \ref{fig:Model-Geo}. The channel between the BS and the ARIS depends on both the 3-D position of the UAV and the orientation of the ARIS. It is expressed as
\begin{equation}
    \resizebox{0.91\hsize}{!}{$%
    \mathbf{H} = \alpha(\mathbf{p})\, \mathbf{a}_{\text{rx},R}\left(\vartheta_{B-R}^{\text{AoA}}, \varphi_{B-R}^{\text{AoA}}, \mathbf{q} \right) \, \mathbf{a}_{\text{tx},B}^\dagger\left(\vartheta_{B-R}^{\text{AoD}}, \varphi_{B-R}^{\text{AoD}}\right),
    $}%
\end{equation}
where $\left(\vartheta_{B-R}^{\text{AoD}}, \varphi_{B-R}^{\text{AoD}}\right)$ are the elevation and azimuth AoDs from the BS, and $\mathbf{a}_{\text{tx},B}$ and $\mathbf{a}_{\text{rx},R}$ are the transmit array response at the BS and the receive array response at the ARIS, respectively. The latter are defined as
\begin{equation}
    \mathbf{a}_{\mathrm{tx},B}(\vartheta_{B-R}^{\mathrm{AoD}},\varphi_{B-R}^{\mathrm{AoD}})
    =
    \left[
    e^{j\,\frac{2\pi}{\lambda}\mathbf{s}(\vartheta_{B-R}^{\mathrm{AoD}},\varphi_{B-R}^{\mathrm{AoD}})\cdot\mathbf{p}_n^{\mathrm{BS}}}
    \right]_{n=1}^{N},
\end{equation}
\begin{equation}
    \resizebox{0.91\hsize}{!}{$%
    \mathbf{a}_{\mathrm{rx},R}(\vartheta_{B-R}^{\mathrm{AoA}},\varphi_{B-R}^{\mathrm{AoA}},\mathbf{q})
    =
    \left[
    e^{j \frac{2\pi}{\lambda}\,
    \Delta \ell_m(\vartheta_{B-R}^{\mathrm{AoA}},\varphi_{B-R}^{\mathrm{AoA}},\mathbf{q})}
    \right]_{m=1}^{M}.
    $}
\end{equation}

Similarly, the channel between the UAV and user $i$, denoted by $\mathbf{g}_{i} \in \mathbb{C}^{1 \times M}$, is given by
\begin{equation}
    \mathbf{g}_{i} = \beta_{i}(\mathbf{p}) \, \mathbf{a}_{\text{tx},R}^\dagger\left(\vartheta_{R-i}^{\text{AoD}}, \varphi_{R-i}^{\text{AoD}}, \mathbf{q} \right),
\end{equation}
where $\mathbf{a}_{\text{tx},R}$ is the RIS transmit array response, expressed as
\begin{equation}
    \mathbf{a}_{\mathrm{tx},R}(\vartheta_{R-i}^{\mathrm{AoD}},\varphi_{R-i}^{\mathrm{AoD}},\mathbf{q})
    =
    \left[
    e^{j \frac{2\pi}{\lambda}\,
    \Delta \ell_{m,i}(\vartheta_{R-i}^{\mathrm{AoD}},\varphi_{R-i}^{\mathrm{AoD}},\mathbf{q})}
    \right]_{m=1}^{M}.
\end{equation}

The complex LoS channel coefficients for the BS-to-ARIS link, $\alpha(\mathbf{p})$, and the ARIS-to-user link, $\beta_{i}(\mathbf{p})$, are given by
\begin{equation}\label{PathLoss_BS_ARIS}
    \alpha(\mathbf{p}) = \frac{\sqrt{l_0}}{\left\| \mathbf{p} - \mathbf{p}^{\text{BS}} \right\|_2} \,
    e^{-j \frac{2\pi \left\| \mathbf{p} - \mathbf{p}^{\text{BS}} \right\|_2}{\lambda}},
\end{equation}
\begin{equation}\label{PathLoss_ARIS_Users}
    \beta_{i}(\mathbf{p}) = \frac{\sqrt{l_0}}{\left\| \mathbf{p} - \mathbf{p}_{i} \right\|_2} \,
    e^{-j \frac{2\pi \left\| \mathbf{p} - \mathbf{p}_{i} \right\|_2}{\lambda}},
\end{equation}
where $l_0$ is the reference path loss at a distance of $1$ meter and $\lambda$ is the carrier wavelength.

The RIS's ability to reflect incoming signals toward a desired direction is strongly influenced by its orientation. As shown above, the orientation influences the channel of communication through its impact on RIS array responses, but it also affects the strength of RIS reflection in a given direction. In particular, the effective reflection strength is direction-dependent and can be quantified by the \textit{aperture gain} function. The unit vectors from the ARIS to the BS and from the ARIS to a user $i$, expressed in the ARIS local frame, are given by

\vspace{-1em}
\begin{equation}
    \mathbf{u}_{B-R}^{\text{local}} = \mathbf{R}(\mathbf{q})^{\top} \frac{\mathbf{p}^{\text{BS}} - \mathbf{p}}{\left\| \mathbf{p}^{\text{BS}} - \mathbf{p} \right\|_2}, 
\end{equation}
\begin{equation}
    \mathbf{u}_{R-i}^{\text{local}} = \mathbf{R}(\mathbf{q})^{\top} \frac{\mathbf{p}_i - \mathbf{p}}{\left\| \mathbf{p}_i - \mathbf{p} \right\|_2}.
\end{equation}

Then, the incident and reflection elevation angles of the dual link $B \to R \to i$ are defined according to \cite{Liu2024ARIS} as
\begin{equation}
    \cos(\varphi_0) = -[\mathbf{u}_{B-R}^{\text{local}}]_3, \quad \cos(\varphi_i) = -[\mathbf{u}_{R-i}^{\text{local}}]_3,
\end{equation}
where $\varphi_0$ and $\varphi_i$ represent the angles between the RIS normal (which coincides with the negative body-fixed $Z_B$ axis) and the directions of arrival and departure, respectively. The \textit{aperture gain} of a user $i$ is then given by
\begin{equation}
    \resizebox{0.89\hsize}{!}{$%
    G^\text{ap}_i=
    \begin{cases}
    \cos(\varphi_0)\cos(\varphi_i), & \cos(\varphi_0)>0,\ \cos(\varphi_i)>0,\\
    0, & \text{otherwise}.
    \end{cases}
    $}%
\end{equation}

This gain factor is in practice always smaller than $1$ and modulates the overall SNR, highlighting the importance of orientation-aware design in ARIS-assisted communication systems.

The SNR of user $i$  depends on both the orientation and position of the ARIS, $\mathbf{q}$ and $\mathbf{p}$, the BS transmit power $P_0$, and the RIS phase shift matrix $\boldsymbol{\Theta} = \operatorname{diag}\left(e^{j \vartheta_{m}}\right)_{m=1}^{M}$ with $\vartheta_{m}$ the phase shift of the $m$-th RIS element. Therefore, under the adopted TDMA protocol, the instantaneous SNR of a scheduled user $i$ is given by
\begin{equation}\label{Eq:SNR_expression}
    \eta_{i} = \frac{P_0 G_A G^{ap}_i \left| \mathbf{g}_{i} \boldsymbol{\Theta} \mathbf{H} \mathbf{f} \right|^2}{\sigma^2},
\end{equation}
where $\mathbf{f}$ is the beamforming vector selected at the BS and $G_A$ is the antenna gain of the BS. Here $\mathbf{H} \in \mathbb{C}^{M\times N}$, $\mathbf{g}_i \in \mathbb{C}^{1 \times M}$, $\boldsymbol{\Theta} \in \mathbb{C}^{M \times M}$, and $\mathbf{f} \in \mathbb{C}^{N \times 1}$, so that $\mathbf{g}_i \boldsymbol{\Theta} \mathbf{H} \mathbf{f}$ is a scalar.


During the scheduled TDMA transmission slot, the instantaneous transmission rate from the BS to user~$i$ is given by
\begin{equation}
    R_{i} =\frac{ B_0}{I} \log_2\left(1 + \eta_{i}\right),
    \label{eq:Rate}
\end{equation}
where $B_0$ is the transmission bandwidth. 



\vspace{-0.5em}
\subsection{UAV dynamics model}
\label{UAV_DM}

The ARIS platform is modeled as a rigid body comprising a multi-rotor UAV and a rigidly mounted RIS panel of negligible mass. Let $\mathbf{p} \in \mathbb{R}^3$, $\mathbf{v} \in \mathbb{R}^3$, and $\boldsymbol{\omega} \in \mathbb{R}^3$ denote the UAV position, linear velocity, and angular velocity, and $\mathbf{q} \in \mathbb{S}^3$ be the unit quaternion representing its orientation.

The UAV dynamics, adopted from~\cite{Leutenegger2016Flying}, are given by
\begin{subequations}\label{eq:dynamics}
    \begin{empheq}[left=\empheqlbrace]{align}
    \dot{\mathbf{p}} &= \mathbf{v}, 
    \label{eq:pos}\\
    \dot{\mathbf{q}} &= \frac{1}{2}\mathbf{q} \circ 
    \begin{bmatrix} 0 \\ \boldsymbol{\omega} \end{bmatrix},
    \label{eq:quat}\\
    m\dot{\mathbf{v}} &= -mg\mathbf{e}_z + u_T \mathbf{R}(\mathbf{q})\mathbf{e}_z,
    \label{eq:translational}\\
    \mathbf{J}\dot{\boldsymbol{\omega}} &=
    -\boldsymbol{\omega} \times (\mathbf{J}\boldsymbol{\omega}) + \boldsymbol{\tau},
    \label{eq:rotational}
    \end{empheq}
\end{subequations}
where $m$ and $\mathbf{J}=\mathrm{diag}(J_x,J_y,J_z)$ are the UAV mass and inertia matrix, $\mathbf{e}_z=[0,\,0,\,1]^\top$, $u_T$ is the total thrust, $\boldsymbol{\tau}=[\tau_x,\,\tau_y,\,\tau_z]^\top$ is the control torque, and $\mathbf{R}(\mathbf{q})\in\mathrm{SO}(3)$ is the rotation matrix associated with $\mathbf{q}$. Equations~\eqref{eq:pos} and \eqref{eq:translational} describe the translational dynamics, and~\eqref{eq:quat}~and~\eqref{eq:rotational}~the rotational motion.

The thrust $u_T$ and torques $\boldsymbol{\tau}$ are generated by the four rotor speeds $\Omega_1,\dots,\Omega_4$, mapped as 
\begin{equation}
\begin{bmatrix}
u_T\\
\tau_x\\
\tau_y\\
\tau_z
\end{bmatrix}
=
\begin{bmatrix}
b_f & b_f & b_f & b_f\\
-\frac{b_f d}{\sqrt{2}} & -\frac{b_f d}{\sqrt{2}} & \frac{b_f d}{\sqrt{2}} & \frac{b_f d}{\sqrt{2}}\\
-\frac{b_f d}{\sqrt{2}} & \frac{b_f d}{\sqrt{2}} & \frac{b_f d}{\sqrt{2}} & -\frac{b_f d}{\sqrt{2}}\\
-b_m & b_m & -b_m & b_m
\end{bmatrix}
\begin{bmatrix}
\Omega_1^2\\
\Omega_2^2\\
\Omega_3^2\\
\Omega_4^2
\end{bmatrix},
\label{eq:rotor_map}
\end{equation}
where $b_f$, $b_m$, and $d$ are the rotor thrust coefficient, drag coefficient, and arm length.

The overall system is compactly written as $\dot{\mathbf{x}} = \mathbf{h}(\mathbf{x},\mathbf{u})$,  with state $\mathbf{x} = [\mathbf{p}^\top,  \mathbf{q}^\top, \mathbf{v}^\top, \boldsymbol{\omega}^\top]^\top \in \mathbb{R}^3 \times \mathbb{S}^3 \times \mathbb{R}^6$, and control input $\mathbf{u} = [\Omega_1,  \Omega_2, \Omega_3, \Omega_4]^\top \in \mathbb{R}^4$.




\section{Problem Statement} 
\label{ProblemFormul}

Our objective is to maximize the network throughput over a finite mission duration by jointly optimizing the UAV motion and the communication variables of the ARIS-assisted system. In particular, we consider the joint design of the BS beamforming vector \(\mathbf{f}\), the RIS phase-shift configuration \(\boldsymbol{\Theta}\), and the UAV control inputs, while explicitly accounting for UAV dynamics, actuation limits, and communication requirements.

The UAV-mounted RIS is required to evolve from a given initial state $\mathbf{x}_I$ to a predefined final state $\mathbf{x}_F$ within a mission time horizon $T$, while ensuring reliable communication with all ground users throughout the flight. The resulting optimization problem tightly couples mobility and communication variables through both the UAV dynamics and the orientation-dependent channel model introduced in Section~\ref{SystAndComModel}.

Formally, the joint optimization problem is formulated as

\vspace{-0.5em}
\begin{small}
    \begin{maxi!}|s|
    {\mathbf{u},\mathbf{f}, \boldsymbol{\Theta}}{\int_{0}^{T} \sum_{i\in \mathcal{I}} R_{i}(t) \, dt}{\label{Problem:Original}}{\label{ObjectiveFunc}}{}
    \addConstraint{\dot{\mathbf{x}}(t) = \mathbf{h}(\mathbf{x}(t), \mathbf{u}(t)) \label{Const:DM}}{}{}
    \addConstraint{\mathbf{x}(0) = \mathbf{x}_I,\ \mathbf{x}(T) = \mathbf{x}_F \label{Const:Init_Final_Positions}}{}{}
    \addConstraint{\mathbf{p}(t) \in \mathcal{A},\ \forall t \in [0, T] \label{Const:Deployed_Area}}{}{}
    \addConstraint{R_{i}(t) \geq R_{\min},\ \forall i \in \mathcal{I}  \label{Const:QoS}}{}{} 
    \addConstraint{\vartheta_{m}(t) \in [0, 2\pi],\ \forall m \in \mathcal{M} \label{Const:Phaseshift}}{}{} 
    \addConstraint{\|\mathbf{v}(t)\|_2 \leq \bar{v},\ \forall t \in [0, T] \label{Const:Velocity}}{}{}
    \addConstraint{ \Omega_s(t) \in [\underline{\Omega},\bar{\Omega}], \forall s \in \{ 1,\dots,4\} \label{Const:rotorspeed}}{}{} 
    \addConstraint{\|\mathbf{f}(t)\|_2 =1,\ \forall t \in [0,T].\label{Const:beamforming}}{}{}
    \end{maxi!}
\end{small}


In~\eqref{ObjectiveFunc}, $R_i(t)$ denotes the achievable data rate of user $i$ at time $t$, defined in~\eqref{eq:Rate}. Constraint~\eqref{Const:DM} enforces the UAV dynamics described in Section~\ref{UAV_DM}, ensuring that the motion is physically consistent with the vehicle’s dynamics. Constraint~\eqref{Const:Init_Final_Positions} specifies the initial and final UAV states, reflecting mission requirements. Constraint~\eqref{Const:Deployed_Area} restricts the UAV position to the admissible flight region $\mathcal{A}$, accounting for operational and safety limitations. The QoS requirement~\eqref{Const:QoS} ensures that each user achieves a minimum instantaneous data rate $R_{\min}$, thereby guaranteeing a baseline level of communication performance. Constraint~\eqref{Const:Phaseshift} defines the feasible range of RIS phase shifts, reflecting the hardware limitations of practical RIS elements. Constraints~\eqref{Const:Velocity} and~\eqref{Const:rotorspeed} enforce the physical limits of the UAV platform, including bounded translational velocity and actuator saturation through rotor speed constraints, ensuring dynamically feasible motion. Finally, constraint~\eqref{Const:beamforming} imposes a unit-norm constraint on the BS beamforming vector, which corresponds to a transmit power normalization at the base station.

Problem~\eqref{Problem:Original} is highly nonconvex due to the nonlinear UAV dynamics, the orientation-dependent channel model, and the multiplicative coupling among mobility, beamforming, and RIS phase-shift variables. Rather than solving \eqref{Problem:Original} globally, in the next section, we develop a decomposition-based approximation that separates communication updates from trajectory generation and tracking, while preserving the dominant coupling between UAV motion and communication performance.




\section{Proposed Solution}
\label{ProposedSolution}

\begin{figure}[tb]
    \centering \includegraphics[width=1\linewidth]{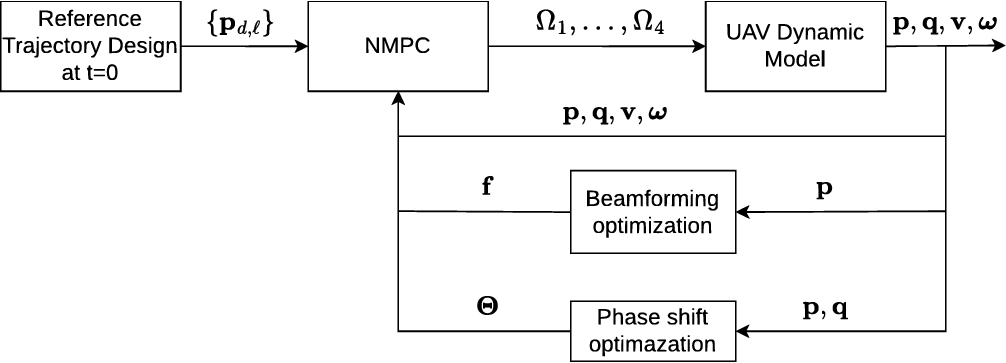}
    \vspace{-1.75em}
    \caption{Block diagram of the proposed sub-optimal control strategy.}
    \label{fig:Block_Diag_optimazation}
    \vspace{-0.75em}
\end{figure}

Solving Problem~\eqref{Problem:Original}  to global optimality is extremely challenging due to the tight coupling between the communication variables (the BS beamforming vector $\mathbf{f}$ and the RIS phase-shift matrix $\boldsymbol{\Theta}$), the UAV dynamics, and the nonlinear rate expressions. Therefore, rather than seeking the globally optimal solution, we develop a tractable solution by adopting a decomposition strategy that separates communication design from trajectory generation and tracking.

This approach is motivated by two main considerations. First, the communication variables evolve on a much faster time scale than the UAV dynamics, allowing beamforming and RIS phase-shift updates to be performed quasi-instantaneously with respect to the NMPC sampling time. Second, when the UAV state is fixed, the structure of the rate expression \eqref{eq:Rate} admits near-optimal closed-form solutions for both the beamformer and the RIS phase shifts under fixed-state assumptions.

As a result, the proposed framework yields a modular architecture, see Fig.~\ref{fig:Block_Diag_optimazation}, in which communication parameters are updated based on the current UAV state, while the UAV trajectory is designed and tracked under the resulting communications-aware constraints. The overall solution consists of four components. First, the BS beamforming vector is computed through the Beamforming optimization block (Section~\ref{Beamforming_Opt}). Second, the RIS phase shifts are updated via the Phase shift optimization block (Section~\ref{Phase_Shift_Opt}). Third, a reference trajectory is generated offline using the Reference Trajectory Design at $t = 0$ block (Section~\ref{OfflineTrajGeneration}). Finally, this reference trajectory is tracked online using the NMPC controller, which incorporates the UAV Dynamic Model and physical limits (Section~\ref{NMPC_Tracking}).


The resulting framework should be interpreted as a low-complexity approximation of the original joint optimization problem. Rather than solving Problem~\eqref{Problem:Original} globally, it preserves the dominant interaction between the communication and control variables by updating the communication parameters as explicit functions of the UAV state. Consequently, although the variables are optimized sequentially rather than jointly, the simplified framework retains the essential structure of the original problem, thereby providing a meaningful low-complexity approximation. Within the proposed decomposition, a fixed horizontal orientation reference is adopted as a stabilization strategy rather than as the communication-optimal attitude. Obtaining the communication-optimal orientation would require jointly optimizing the UAV trajectory and attitude while accounting for their coupled dynamics, resulting in a significantly more complex optimization problem.



\vspace{-0.25em}
\subsection{Beamforming optimization}
\label{Beamforming_Opt}

We optimize the beamforming vector $\mathbf{f}$ to maximize the total instantaneous throughput at time $t$, while fixing the UAV state $\mathbf{x}$ and the RIS phase-shift matrix $\boldsymbol{\Theta}$. The resulting optimization problem is formulated as \eqref{Problem:BF_1}.

\vspace{-1em}
\begin{small}
\begin{maxi!}|s|
{\mathbf{f}}{\sum_{i\in \mathcal{I}} R_{i}(t) }{\label{Problem:BF_1}}{}{}
\addConstraint{\eqref{Const:beamforming}.}{}{}
\end{maxi!}
\end{small}

At this stage, we adopt the Maximum Ratio Transmission (MRT) toward the BS-to-RIS channel as a low-complexity surrogate while the constraint \eqref{Const:QoS} is deferred to the NMPC tracking phase (Section \ref{NMPC_Tracking}). For the fixed-state setting considered here, this choice maximizes the power transferred from the BS array toward the RIS and provides a practical beamforming update that is consistent with the subsequent RIS reflection optimization. To achieve this, at each time $t$, the BS uses the following beamforming vector
\begin{equation}
    \mathbf{f} = \frac{\mathbf{a}_{\text{tx},B}\left(\vartheta_{B-R}^{\text{AoD}}, \varphi_{B-R}^{\text{AoD}}\right)}
    {\left\|\mathbf{a}_{\text{tx},B}\left(\vartheta_{B-R}^{\text{AoD}}, \varphi_{B-R}^{\text{AoD}}\right) \right\|_2}.
\end{equation}

This strategy is optimal for maximizing SNR at the RIS~\cite{jeon2022energy}.



\subsection{Phase shift optimization}
\label{Phase_Shift_Opt}

We optimize the RIS phase-shift matrix $\boldsymbol{\Theta}$ at time $t$ to maximize the instantaneous throughput, while fixing the UAV state $\mathbf{x}$ and the beamforming vector $\mathbf{f}$. The optimization problem is formulated as follows

\vspace{-1em}
\begin{small}
\begin{maxi!}|s|
{\boldsymbol{\Theta}}{ \sum_{i\in \mathcal{I}}  R_{i}(t)}{}{}{\label{Problem:PS_1}}
\addConstraint{\eqref{Const:Phaseshift}. }{}{}
\end{maxi!}
\end{small}

The QoS constraint is disregarded in this step. The single-user equivalent formulation enables a closed-form phase-shift expression that can be evaluated efficiently and reused to predict future channel states. In contrast, performing multi-user phase-shift optimization with fixed beamforming offers no closed-form solution and would require designing an iterative method with non-negligible computational complexity.


To reduce complexity, we define a virtual user located at the barycenter of the users' positions, denoted by $\mathbf{p}^\text{v}$, and apply the single-user solution to this representative node. This approximation is appropriate for clustered users, as considered in this work, although it may become inefficient if the users are widely dispersed. The simulation results confirm the effectiveness of this strategy under the considered setup. According to the single-user analysis in~\cite{Liu2024ARIS}, the optimal phase shifts are derived as
\begin{equation}
\begin{split}
\vartheta_{m} 
&= \angle \left( \left( \mathbf{a}_{\text{tx},R}\left(
    \vartheta_{R-v}^{\text{AoD}}, 
    \varphi_{R-v}^{\text{AoD}}, 
    \mathbf{q}
\right) \right)_m \right) \\
&\quad - \angle \left( \left( \mathbf{a}_{\text{rx},R}\left(
    \vartheta_{B-R}^{\text{AoA}}, 
    \varphi_{B-R}^{\text{AoA}}, 
    \mathbf{q}
\right) \right)_m \right), \forall m \in \mathcal{M},
\end{split}
\end{equation}
where $\left(\vartheta_{R-v}^{\text{AoD}}, \varphi_{R-v}^{\text{AoD}}\right)$ are the angles from the RIS to the virtual user. 

This phase-shift update is optimal for the adopted single-user surrogate centered at the user barycenter. Therefore, within the overall framework, RIS phase design should be interpreted as a low-complexity approximation of the original multi-user phase-shift optimization problem.



\subsection{Offline trajectory generation}
\label{OfflineTrajGeneration}

\begin{algorithm}[tb]
\caption{Reference Trajectory Design}
\label{alg:knn_dijkstra}
\begin{algorithmic}[1]

\Statex \textbf{Input:} $\mathcal{A}$, $p_z= h^R$, initial position of the UAV $\mathbf{p}^{\text{init}}$, final position of the UAV $\mathbf{p}^{\text{fin}}$, neighborhood size $E \geq 2$, total number of vertices $L$, flight duration $T$, and sampling time $T_s$.
\Statex \textbf{Output:} Communication-guided reference trajectory $\mathcal{P}.$

\State Randomly generate $L$ candidate nodes $\mathcal{V} = \{(p_x^l, p_y^l)\}_{l=1}^{L}$ within the rectangular area in $\mathcal{A}$ at altitude $p_z$, by ensuring that $\mathbf{p}^{\text{init}}$ and $\mathbf{p}^{\text{fin}}$ are included in $\mathcal{V}$.

\State Compute the cost of each node as $C_l$ for all $l \in \{1, \dots, L\}$ from \eqref{Eq:Cost_Computation} .

\State Construct a graph $\mathcal{G}$ over the set of vertices $\mathcal{V}$, where each vertex is connected to its $E$ nearest neighbors based on Euclidean distance, forming a weighted graph with edge weights $\{w_{i_0 \to j_0}\}$.\label{ln:knn}

\ForAll{$i_0 \in \{1, \dots, L\}$}
    \ForAll{$j_0$ in the $E$ nearest neighbors of $i_0$}
        \State $w_{i_0 \to j_0} \gets C_{j_0}$
    \EndFor
\EndFor

\State Run \textbf{Dijkstra's algorithm} on the weighted graph $(\mathcal{V}, w)$ from node $\mathbf{p}^{\text{init}}$ to node $\mathbf{p}^{\text{fin}}$ to compute the shortest path $\mathcal{P}$.

\State Let $N_T = \lfloor T/T_s\rfloor + 1$ be the number of reference samples. Interpolate the waypoints along the path $\mathcal{P}$ to generate a sequence of $N_T$ equally spaced reference positions. The resulting sequence constitutes the reference trajectory.

\end{algorithmic}
\end{algorithm}

In this subsection, we compute a communication-guided spatial reference path between the initial and final positions. In the offline graph-based planner, the UAV altitude is kept constant and only the horizontal motion is optimized. In the world frame $\mathcal{F}_W$, the fixed planning altitude is set to $p_z = h^R$. The graph-search planner does not directly optimize the full orientation-aware sum-rate; instead, it uses a lightweight distance-based surrogate to identify communication-favorable regions at low computational cost. Orientation-dependent effects are then handled online by the NMPC and the state-dependent communication updates.

To generate the reference 2D trajectory, we first sample candidate waypoints in the spatial deployment area with respect to the $x-$ and $y-$ coordinates while keeping the altitude fixed. We then construct a weighted graph over the resulting candidate positions and search for a path from the initial to the final location that minimizes a communications-aware cost.

In principle, the edge weight between two nodes (the cost) should be defined as the negative network sum-rate evaluated at the arrival position. However, this requires computing the full communication model for every edge in the graph, leading to a complexity of $\mathcal{O}(LINM)$, where $L$ is the number of vertices, $I$ is the total number of users, $N$ is the number of antenna elements, and $M$ is the total number of RIS elements. To reduce this computational burden, we introduce a simplified cost function that approximates the communication quality of each vertex without explicitly evaluating the full data rate expression. The cost at a node $l$ is defined as follows
\begin{equation}\label{Eq:Cost_Computation}
    C_l = (\|\mathbf{p}^l - \mathbf{p}^\text{v}\|_2 \cdot \|\mathbf{p}^l - \mathbf{p}^{\text{BS}}\|_2)^\gamma,
\end{equation}
where $\mathbf{p}^l = [p_x^l, p_y^l, h^R]^\top$ denotes the 3D position of the vertex $l$ in the graph $\mathcal{G}$ and $\gamma\in \mathbb{R}_{+}$ denotes a factor to enforce the importance of communication in the design of the reference trajectory. This cost corresponds to the product of the distances from the vertex to $\mathbf{p}^\text{v}$, the virtual user position, and from the vertex to the BS position $\mathbf{p}^{\text{BS}}$. This is motivated by the spatial clustering of the users. In the general case, the cost at node $l$ is $ C_l = ( \|\mathbf{p}^l - \mathbf{p}^{\text{BS}}\|_2 \times \prod_{i=1}^I \|\mathbf{p}^l - \mathbf{p}_i\|_2 )^\gamma$. A low cost value indicates that the UAV is close to both the BS and the users when positioned at $\mathbf{p}^l$. This is consistent with the structure of the path loss functions $\alpha(\mathbf{p})$ and $\beta_i(\mathbf{p})$, respectively, in \eqref{PathLoss_BS_ARIS} and \eqref{PathLoss_ARIS_Users}, where shorter distances reduce path loss. Moreover, these distances determine both the amplitude attenuation through the inverse-distance path-loss terms and the propagation phase through the exponential terms of the channel coefficients. Consequently, minimizing the proposed cost favors locations that simultaneously reduce path-loss attenuation. The complexity of the cost computation is reduced to $\mathcal{O}(L)$. 

The 2D reference trajectory is generated using Algorithm~\ref{alg:knn_dijkstra}. First, the spatial domain is sampled to construct a set of candidate nodes that includes both the initial and terminal positions. A weighted graph is then built by connecting each node to its  $E$ nearest neighbors based on Euclidean distance. The edge weights are defined from the cost associated with the destination node, and Dijkstra’s algorithm is applied to compute the minimum-cost path on the resulting graph. The discrete path is subsequently densified to match the NMPC prediction grid, yielding the reference trajectory used for tracking. Alternative graph-search methods such as A$^\star$ could also be used. However, for the moderate graph sizes considered here, its computational advantage over Dijkstra’s algorithm is limited. Sampling-based planners such as RRT$^\star$ are less suitable in this setting, since the problem is formulated on a finite graph rather than as a continuous high-dimensional motion-planning problem. Under this formulation, Dijkstra’s algorithm provides a deterministic and graph-optimal solution on the constructed roadmap.



\begin{table}[tb]
    \centering
    \scriptsize
    \caption{UAV Dynamic Model Parameters and NMPC Weights}
    \vspace{-1em}
    \label{Tab_NMPC_DM_Params}
    \renewcommand{\arraystretch}{1.1}
    \begin{tabular}{l c c}
    \hline
    \textbf{Parameter} & \textbf{Value} & \textbf{Unit } \\
    \hline
    \multicolumn{3}{c}{\textbf{UAV Dynamic Model}} \\
    \hline
    Mass & $m = 1.042$ & kg \\
    Gravity & $g = 9.81$ & m/s$^2$ \\
    Arm length & $d = 0.23$ & m \\
    Inertia matrix & $J = \mathrm{diag}(0.015, 0.015, 0.070)$ & kg$\cdot$m$^2$ \\
    Thrust coefficient & $b_f = 5.95 \times 10^{-4}$ & N/Hz$^2$ \\
    Moment coefficient & $b_m = 1\times 10^{-5}$ & N$\cdot$m/Hz$^2$ \\
    Rotor speed limits & $\underline{\Omega}=16,\; \bar{\Omega}=100$ & Hz \\
    Maximum velocity & $\bar{v} = 17$ & m/s \\
    \hline
    \multicolumn{3}{c}{\textbf{NMPC}} \\
    \hline
    Sampling time & $T_s = 80$ & ms \\
    Prediction horizon & $N_p = 15$ & steps \\
    Position weight & $\mathbf{Q}_p= \mathbf{Q}_{p,N_p} =\mathrm{diag}(10,\,10,\,12)$ & -- \\
    Velocity weight & $\mathbf{Q}_v= \mathbf{Q}_{v,N_p}  = 1.1\,\mathrm{diag}(1,\,1,\,1)$ & -- \\
    Angular velocity weight & $\mathbf{Q}_\omega= \mathbf{Q}_{\omega,N_p}  = 30\,\mathrm{diag}(1,\,1,\,1)$ & -- \\
    Quaternion weight & $Q_{q} =Q_{q,N_p}= 10$ & -- \\
    Slack variable weight & $\mathbf{Q}_{sv} = 2 \times 10^{-12}\, \mathrm{diag}(1,\,\ldots,\,1)$ & -- \\
    \hline
    \end{tabular}
\end{table}


\subsection{NMPC-based trajectory tracking}\label{NMPC_Tracking}

Given the reference trajectory generated in Section~\ref{OfflineTrajGeneration} and the communication parameters computed at the current sampling instant, the NMPC solves a finite-horizon OCP to determine feasible control inputs that track the reference while satisfying the UAV dynamics and physical constraints. The beamforming vector and RIS phase-shift matrix are computed from the current UAV state before the NMPC problem is solved, and are held constant over the prediction horizon. They are recomputed at the next sampling instant using the newly measured UAV state.

\subsubsection{Discrete-time formulation} The system evolves over a finite prediction horizon of $N_p$ steps with sampling period $T_s > 0$, and corresponding real time $t_k = kT_s$, where the subscript $k$ is an integer number. For time-indexed sequences, the subscript $\ell \in \{0, \dots, N_p\}$ denotes the value of the sequence at step $\ell$ (e.g., $\mathbf{x}_\ell$, $\mathbf{p}_\ell$, $\mathbf{v}_\ell$). This discretization embeds the continuous-time dynamics of Section \ref{UAV_DM} into the NMPC framework. The communication layer operates at a faster time scale than the NMPC. Accordingly, each sampling interval $T_s$ is uniformly divided into $I$-TDMA transmission slots, where $I$ is the number of users (for a sampling time of $T_s=0.08$~s, corresponding to an NMPC update rate of $12.5$~Hz, and a total of $I=5$ users, the corresponding TDMA update rate is $62.5$~Hz). During each sub-slot, only one user is served using the full system bandwidth $B_0$.

\subsubsection{Objective function} The tracking objective is to drive the UAV state toward a prescribed reference trajectory. The NMPC state is defined as $\mathbf{x} = [\mathbf{p}^\top, \mathbf{q}^\top, \mathbf{v}^\top, \boldsymbol{\omega}^\top]^\top$. At prediction step $\ell$, the corresponding desired state is $\mathbf{x}_{d,\ell} =  [\mathbf{p}_{d,\ell}^\top, \mathbf{q}_{d,\ell}^\top, \mathbf{v}_{d,\ell}^\top, \boldsymbol{\omega}_{d,\ell}^\top]^\top$, where $\mathbf{p}_{d,\ell}$ is provided by the reference trajectory generated in Section \ref{OfflineTrajGeneration}.

Position, velocity, and angular-velocity tracking errors are penalized through quadratic terms weighted by positive-definite diagonal matrices. Since quaternion discrepancies cannot be meaningfully measured using a Euclidean norm, the attitude tracking error is instead quantified through the geodesic quaternion distance $d_q(\mathbf{q}_\ell,\mathbf{q}_{d,\ell})$, following the definition in~\cite{Dmytruk2022NMPC}.

The instantaneous rate constraint $R_i(\mathbf{x}_\ell)\geq R_{\min}$ in~\eqref{Const:QoS} may become temporarily infeasible in the presence of disturbances or model mismatch. To preserve feasibility of the NMPC problem, nonnegative slack variables $s_{i,\ell}\geq 0$ are introduced for each user and prediction step. These slack variables are penalized in the cost function through the quadratic term $\|\mathbf{s}_\ell\|_{\mathbf{Q}_{sv}}^2,$ where $\mathbf{s}_\ell=[s_{1,\ell},\dots,s_{I,\ell}]^\top$ and $\mathbf{Q}_{sv}$ is a diagonal matrix with positive weights. This formulation encourages satisfaction of the QoS constraints whenever feasible, while allowing limited constraint relaxation when necessary to maintain robustness and recursive feasibility.

The stage cost at step $\ell$ is defined as follows
\begin{equation}\label{NMPC_Tracking_stage_objective}
\resizebox{0.88\hsize}{!}{$%
\begin{aligned}
e_\ell &= 
\|\mathbf{p}_{d,\ell} - \mathbf{p}_\ell\|_{\mathbf{Q}_p}^2 
+ Q_{q}\mathrm{d}_q(\mathbf{q}_\ell, \mathbf{q}_{d,\ell})^2
\\
&\quad+ \|\mathbf{v}_{d,\ell} - \mathbf{v}_\ell\|_{\mathbf{Q}_v}^2 
+ \|\boldsymbol{\omega}_{d,\ell} - \boldsymbol{\omega}_\ell\|_{\mathbf{Q}_{\omega}}^2
+ \|\mathbf{s}_\ell\|_{\mathbf{Q}_{sv}}^2 ,
\end{aligned}
$}%
\end{equation}
where $\mathbf{Q}_{p}$ is a positive-definite diagonal tracking weight, $Q_{q}$ is a positive scalar tracking weight, while $\mathbf{Q}_{v}$, $\mathbf{Q}_{\omega}$ and $\mathbf{Q}_{sv}$ are positive semi-definite diagonal regularization weights.

To enhance convergence to the terminal equilibrium and support closed-loop stability, a quadratic terminal cost is incorporated as follows
\begin{equation}\label{NMPC_Final_stage_objective}
\resizebox{0.88\hsize}{!}{$%
\begin{aligned}
e_{N_p} &= 
\|\mathbf{p}_{d,{N_p}} - \mathbf{p}_{N_p}\|_{\mathbf{Q}_{p,N_p}}^2 
+ Q_{q,N_p}\mathrm{d}_q(\mathbf{q}_{N_p}, \mathbf{q}_{d,N_p})^2
 \\
&\quad
+ \|\mathbf{v}_{d,N_p} - \mathbf{v}_{N_p}\|_{\mathbf{Q}_{v,N_p}}^2+ \|\boldsymbol{\omega}_{d,N_p} - \boldsymbol{\omega}_{N_p}\|_{\mathbf{Q}_{\omega,N_p}}^2,
\end{aligned}
$}%
\end{equation}
where $\mathbf{Q}_{p,N_p}$, $\mathbf{Q}_{v,N_p}$ and $\mathbf{Q}_{\omega,N_p}$ are positive semi-definite diagonal terminal weight matrices, and $Q_{q,N_p}$ is a positive scalar terminal weight.

\subsubsection{Optimal Control Problem}
At time $t_k$, the NMPC solves the following finite-horizon optimal control problem over a prediction horizon of $N_p$ steps:

\vspace*{-1.25em}
\begin{small}
    \begin{mini!}|s|
    {\mathbf{u},\,\mathbf{s}}{\sum_{\ell=0}^{N_p-1}
    e_\ell + e_{N_p} }{}{}{\label{eq:opt_problem_OCP}}
    \addConstraint{\mathbf{x}_0 = \mathbf{x}(t_k)\label{Const:init_pos_OCP}}{}{}
    \addConstraint{\mathbf{x}_{\ell+1} = \mathbf{h}_d(\mathbf{x}_\ell, \mathbf{u}_\ell),\ \ell \in \{0, \dots, N_p-1\}\label{Const:DM_OCP}}{}{}
    \addConstraint{\mathbf{p}_\ell \in \mathcal{A}, \ \ell \in \{0,\ldots, N_p\}\label{Const:Deployed_Area_OCP}}{}{}
    \addConstraint{\|\mathbf{v}_\ell\|_2 \leq \bar{v}, \ \ell \in \{0,\ldots, N_p\} \label{Const:Velocity_OCP}}{}{}
    \addConstraint{ \Omega_{s,\ell} \in [\underline{\Omega},\bar{\Omega}], 
    \forall s \in \{ 1,2,3,4\}, \  \ell \in \{0,\ldots,N_p\} \label{Const:rotorspeed_OCP}}{}{}
    \addConstraint{R_{i}(\mathbf{x}_\ell) + s_{i,\ell} \geq R_{\min},\  i \in \mathcal{I},\ \ell \in \{0,\ldots, N_p\} \label{Const:QoS_OCP}}{}{}
    \addConstraint{s_{i,\ell} \geq 0,\  i \in \mathcal{I}, \ \ell \in \{0,\ldots, N_p\}. \label{Const:SlackVar_OCP}}{}{}
    \end{mini!}
\end{small}

In \eqref{eq:opt_problem_OCP}, the objective sums the stage-wise tracking costs over the prediction horizon and a terminal tracking cost.
Constraint~\eqref{Const:init_pos_OCP} sets the measured state at the current sampling instant $t_k$ as the initial condition of the prediction model. Constraint~\eqref{Const:DM_OCP} enforces the discrete-time system dynamics $\mathbf{x}_{\ell+1}=\mathbf{h}_d(\mathbf{x}_\ell,\mathbf{u}_\ell)$ at each prediction step. Constraints~\eqref{Const:Deployed_Area_OCP} and~\eqref{Const:Velocity_OCP} keep the predicted trajectory within the admissible deployment region $\mathcal{A}$ and within the maximum-velocity bound, respectively. Constraint~\eqref{Const:rotorspeed_OCP} enforces actuator limits by restricting each rotor speed to $[\underline{\Omega},\bar{\Omega}]$.
Constraint~\eqref{Const:QoS_OCP} is a softened version of the instantaneous rate requirement in~\eqref{Const:QoS}. As a hard constraint,~\eqref{Const:QoS} can become infeasible under model mismatch or disturbances; the slack variables relax it so that the OCP itself always admits a solution, while their quadratic penalization keeps QoS violations as small as possible.


\subsubsection{NMPC online complexity analysis}\label{ComplexityAnalysis} In the proposed framework, at each sampling time, the NMPC formulation takes the form of a nonlinear optimization problem, which is linearized and transformed into a sequence of quadratic programs~\cite{Chen2019matmpc}. Solving a nonlinear program requires cubic computational effort, expressed as $\mathcal{O}\big(I_{0}(n_z + n_a)^3\big)$, where $n_z$ is the total number of decision variables, $n_a$ the total number of active constraints, and $I_{0}$ the number of solver iterations~\cite{Ferreau2014qpOASES}. The computational complexity discussed above characterizes the optimization problem itself. The actual execution time depends on the target onboard hardware, the numerical solver, and implementation-specific details. To complement the theoretical complexity analysis with a practical assessment, the execution time of the proposed framework is reported and discussed in Section~\ref{sec:dynamics}.






\section{Simulation Results}
\label{SimResults}

\begin{figure}[t]
    \centering
    \includegraphics[width=0.9\columnwidth]{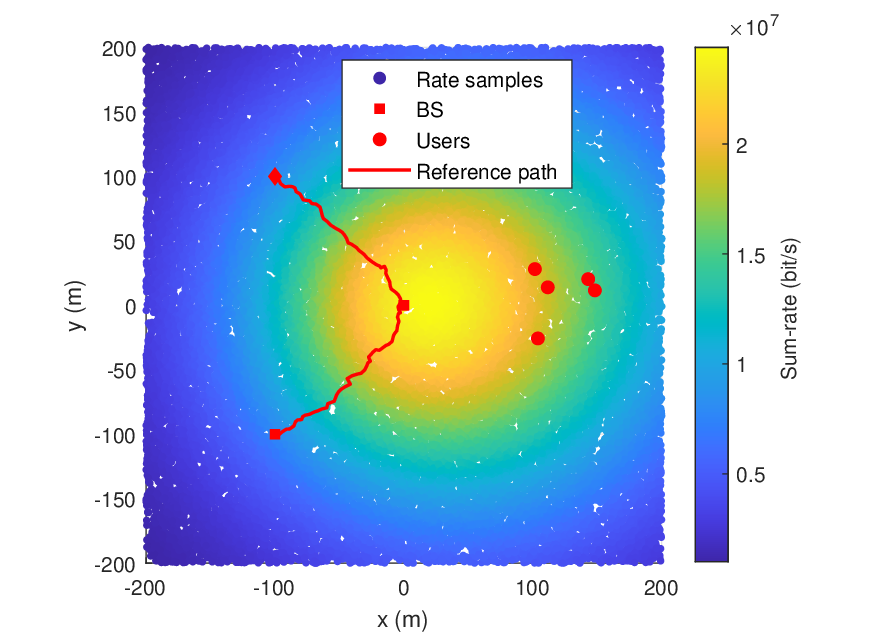}
    \caption{Reference trajectory over the communication-aware cost map, showing the selected path toward favorable regions.}
    \label{fig:Reference_Trajectory}
\end{figure}

\begin{figure}[t]
    \centering
    \includegraphics[width=0.9\columnwidth]{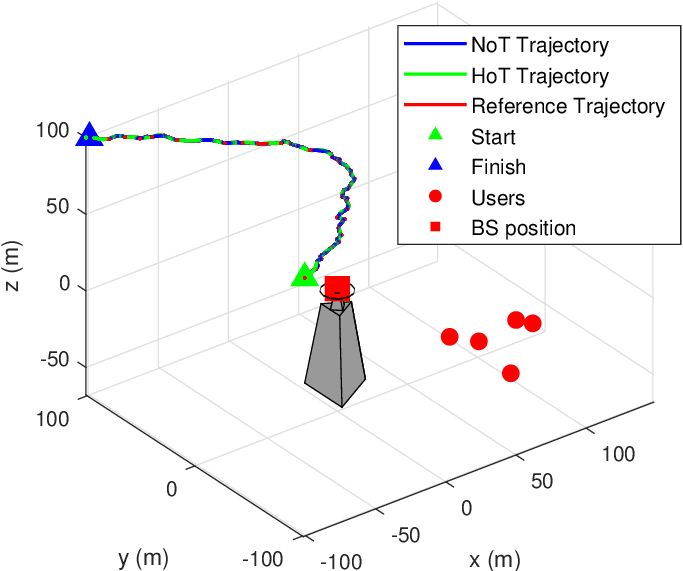}
    \caption{3D UAV trajectory tracking, where curves represent reference, HoT, and NoT paths, illustrating overall tracking behavior.}
    \label{fig:3D_Trajectory}
\end{figure}

\begin{figure}[t]
    \centering
    \vspace{0.2cm}
    \includegraphics[width=0.9\columnwidth]{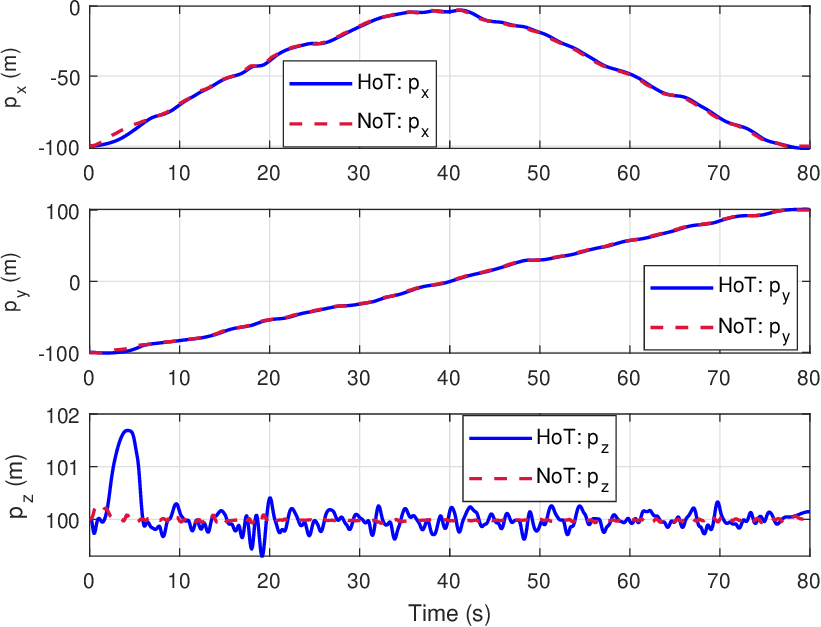}
    \caption{UAV position components over time for HoT and NoT, showing tracking performance and altitude variations.}
    \label{fig:3DPosVsTime}
\end{figure}

\begin{figure}[t]
    \centering
    \vspace{0.45cm}
    \includegraphics[width=0.9\columnwidth]{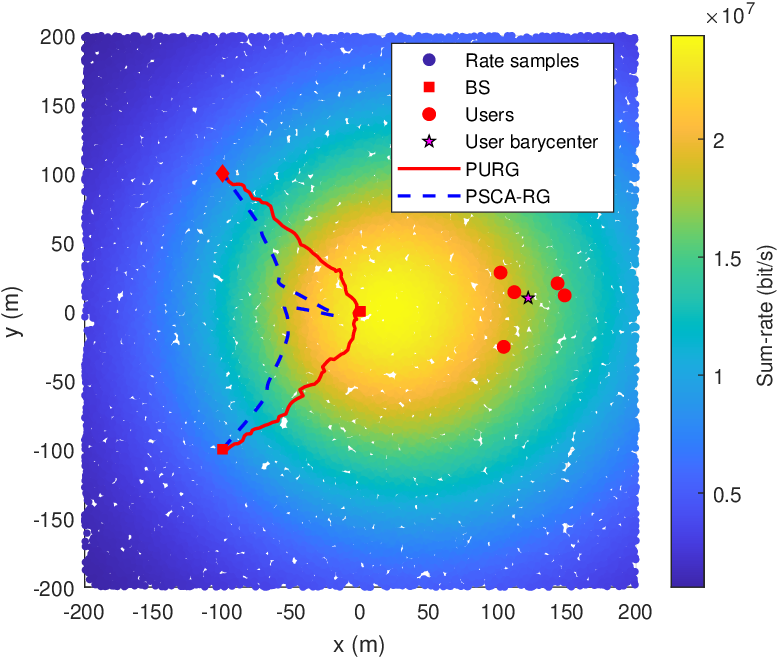}
    \caption{Comparison of the proposed PURG trajectory with the PSCA-RG benchmark in the 2D spatial plane.}
    \label{fig:Alg1_Vs_PSCA:Ref_Traj}
\end{figure}

\begin{figure}[t]
    \centering
    \vspace{0.2cm}
    \includegraphics[width=0.9\columnwidth]{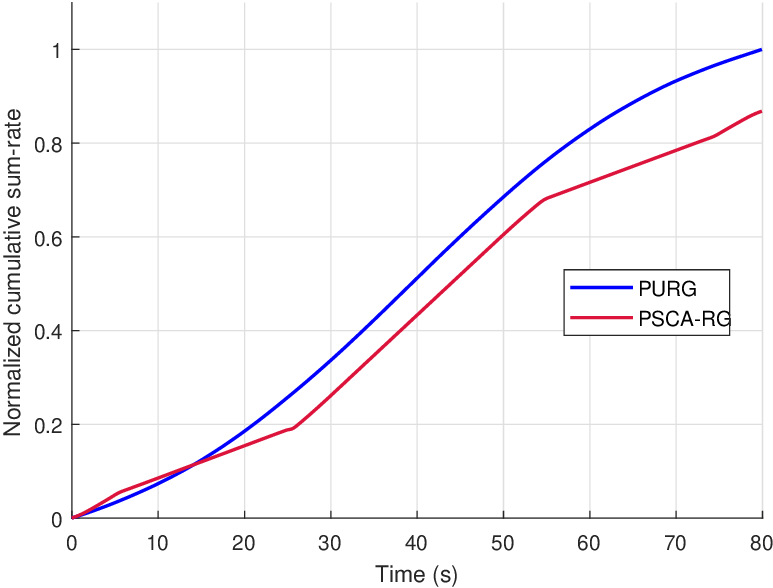}
    
    \caption{Normalized cumulative sum-rate achieved by the reference trajectories generated by PURG and PSCA-RG over the mission duration. The curves are normalized by the maximum value attained by the PURG scheme.}
    \label{fig:Alg1_Vs_PSCA:Cum_Data_Rate}
\end{figure}

\begin{figure}[t]
    \centering
    \vspace{0cm}
    \includegraphics[width=0.9\columnwidth]{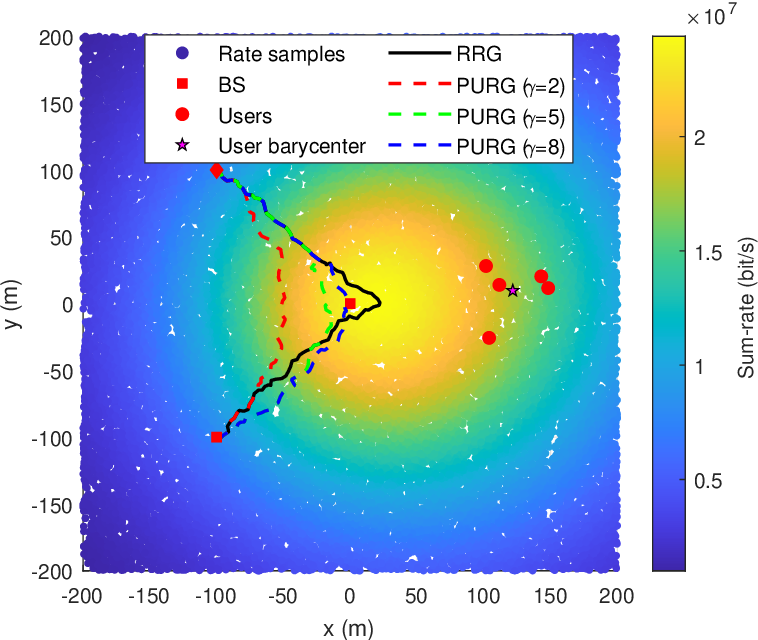}
 
    \caption{Sensitivity analysis of the PURG trajectory for different values of the proxy utility exponent $\gamma$ compared to the RRG trajectory.}
    \vspace{-0.5cm}
    \label{fig:Alg1_PU_Vs_Rate}
\end{figure}

\begin{figure}[t]
    \centering
    \vspace{0.8cm}
    \includegraphics[width=0.9\columnwidth]{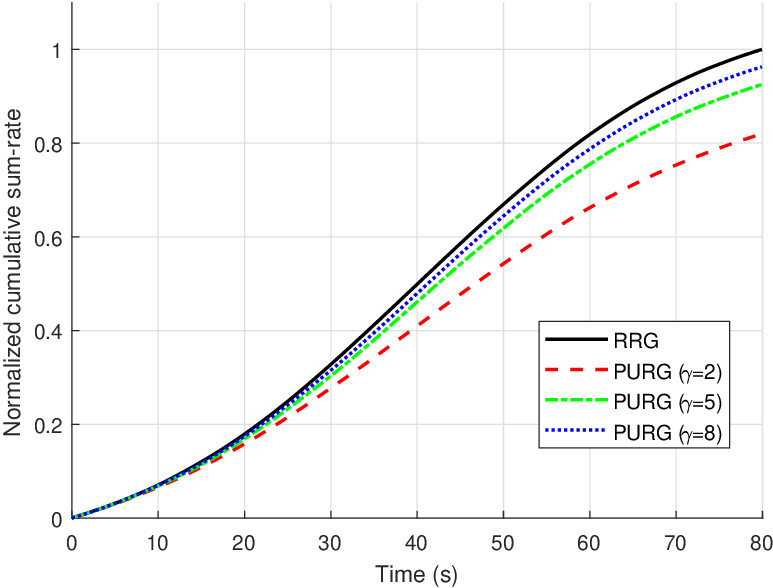}
    
    \caption{Normalized cumulative sum-rate for different $\gamma$ values, showing the convergence of PURG toward the RRG as $\gamma$ increases. All curves are normalized by the maximum cumulative sum-rate attained over all methods.}
    \label{fig:CumRate_Alg1_PU_Vs_Rate}
\end{figure}

\begin{figure}[t]
    \centering
     \vspace{-0.27cm}
    \includegraphics[width=0.9\columnwidth]{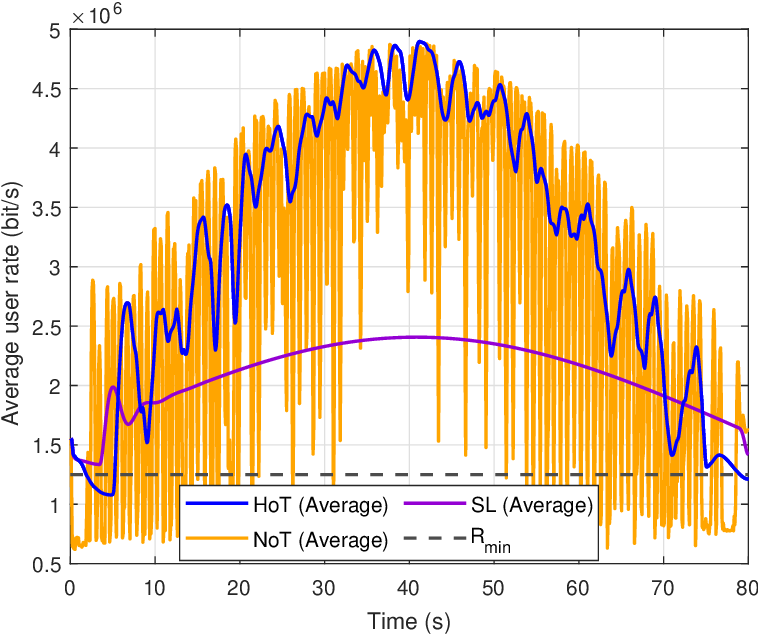}
   
    \caption{Average user data rate versus time, comparing HoT, NoT, and SL schemes, highlighting improved stability and QoS compliance.}
    \label{fig:Users_Data_Rate}
\end{figure}

\begin{figure}[t]
    \centering
    \includegraphics[width=0.9\columnwidth]{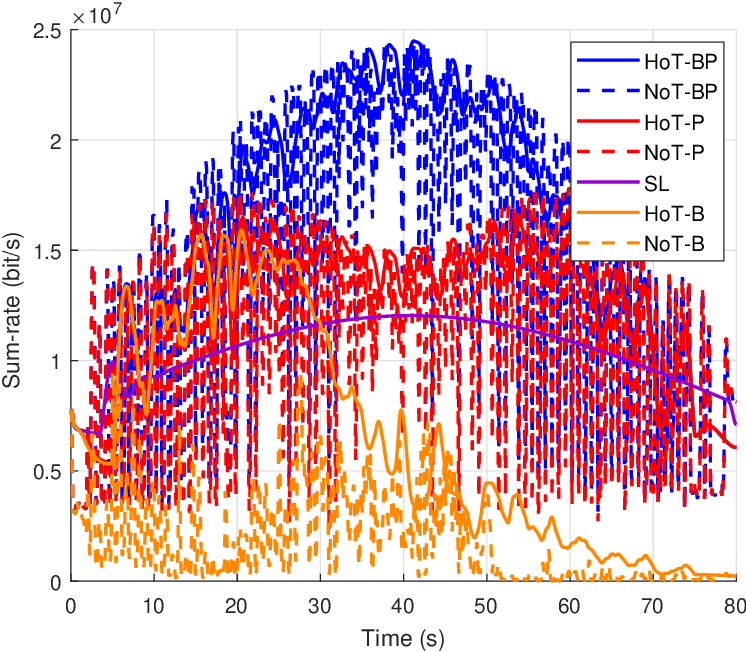}
    
    \caption{Sum-rate over time for different schemes, showing the benefit of joint optimization and orientation tracking.}
    
    \label{fig:Network_Data_Rate}
\end{figure}

\begin{figure}[t]
    \centering
   \vspace{0.02cm}
    \includegraphics[width=0.9\columnwidth]{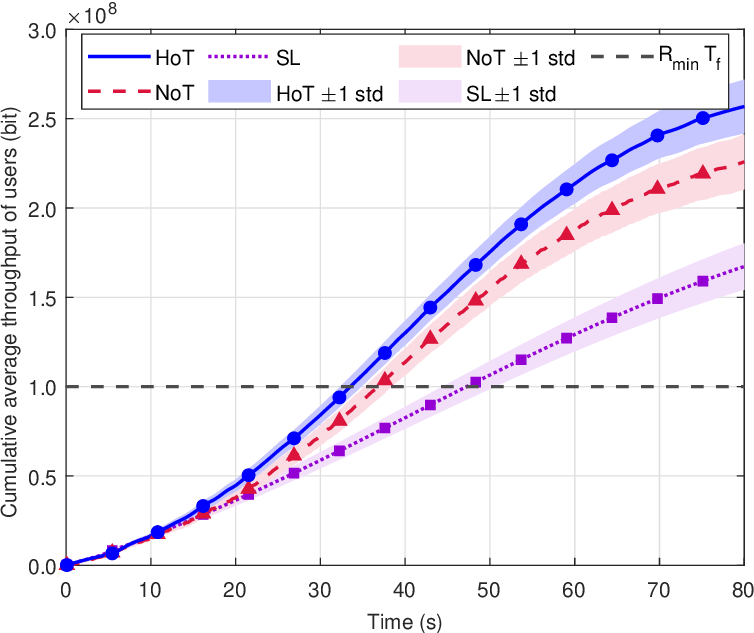}
    
    \caption{Average cumulative throughput per user over time, illustrating the benefits of orientation tracking and communication-aware trajectory design.}
    \label{fig:Users_Throughput}
\end{figure}

\begin{figure}[t]
    \centering
    \vspace{-0.1cm}
    \includegraphics[width=0.9\columnwidth]{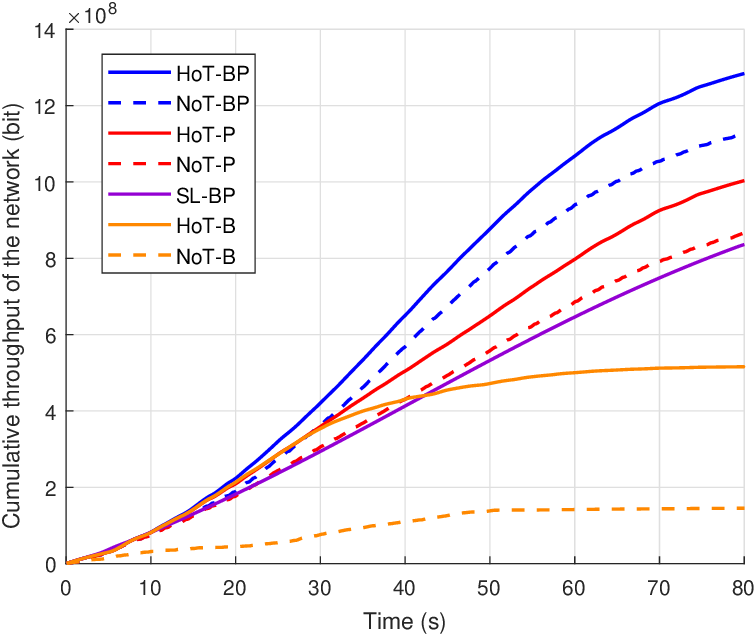}
    
    \caption{Cumulative throughput, showing improved throughput for orientation-aware and joint optimization strategies.}
    \label{fig:Network_Throughput}
\end{figure}

\begin{figure}[t]
    \centering
    \includegraphics[width=0.9\columnwidth]{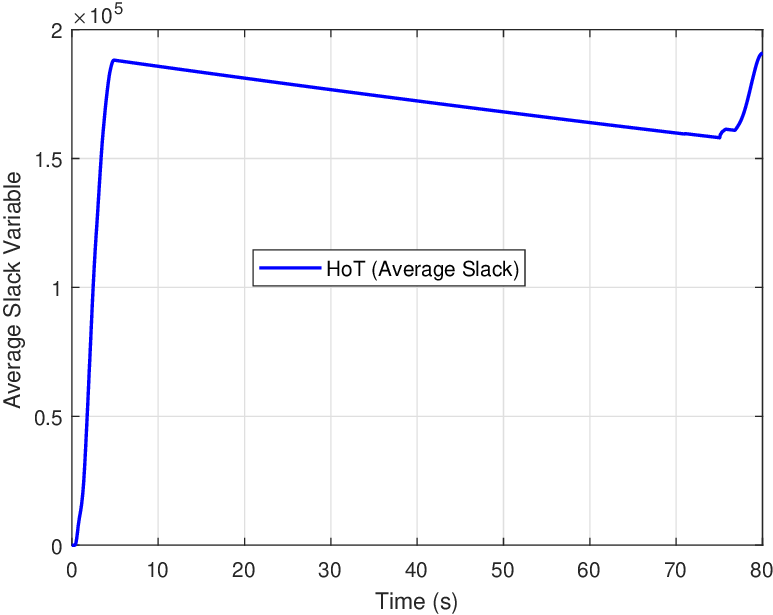}
    \caption{Evolution of the average QoS slack variable over time for the HoT scheme. The QoS slack is averaged over all users.}
    \label{fig:Average_Slack}
\end{figure}

\begin{figure}[t]
    \centering
    \vspace{-0.1cm}
    \includegraphics[width=0.9\columnwidth]{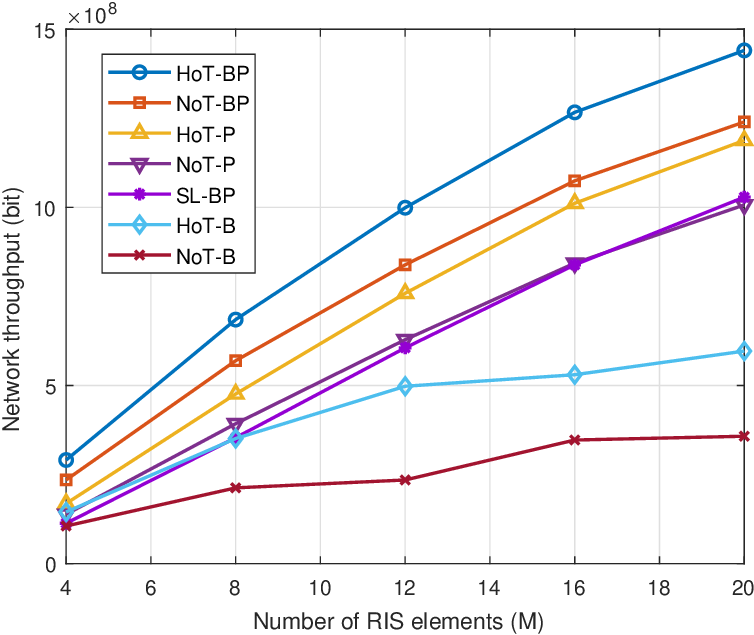}
    
    \caption{Network throughput versus number of RIS elements, demonstrating scalability and performance improvement trends.}
    \label{fig:Scale_NRIS}
\end{figure}

\begin{figure}[t]
    \centering
    \vspace{-0.22cm}
    \includegraphics[width=0.9\columnwidth]{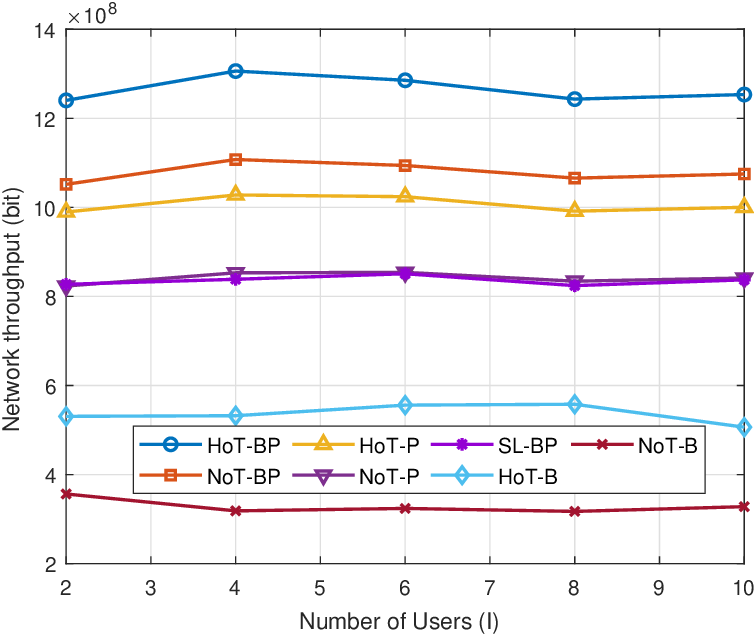}
    \caption{Network throughput versus number of users, highlighting efficiency of the proposed approach.}
    \label{fig:Scale_NUsers}
\end{figure} 

\begin{figure}[t]
    \centering
    \includegraphics[width=0.9\columnwidth]{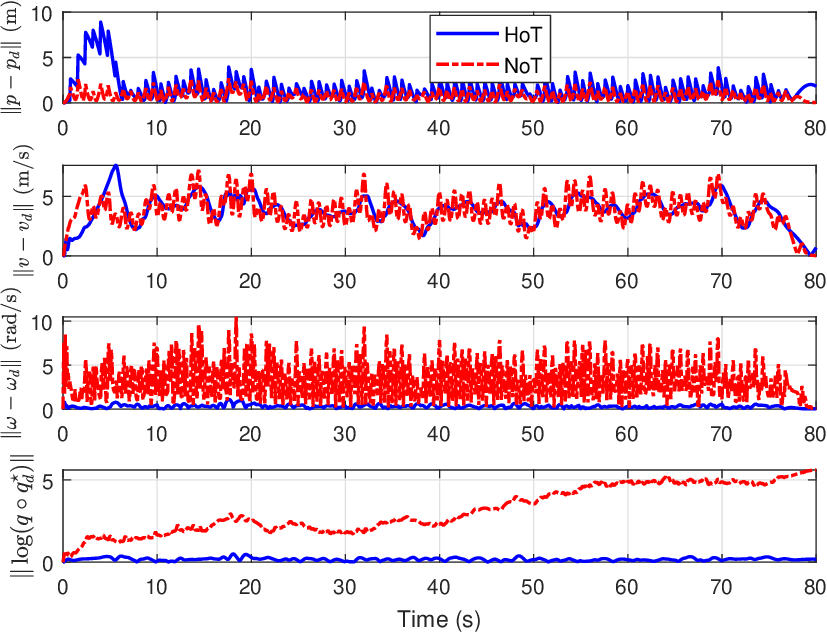}
    \caption{Tracking errors over time, comparing HoT and NoT, and illustrating trade-offs between position and orientation tracking.}
    \label{fig:Tracking_errors}
\end{figure}

\begin{figure}[t]
    \centering
    \includegraphics[width=0.9\columnwidth]{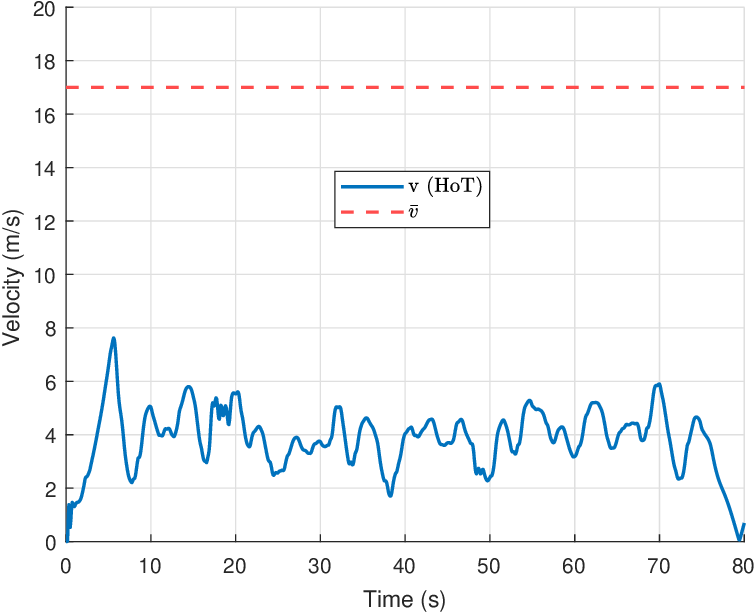}
    \caption{UAV velocity profile over time, showing smooth evolution within imposed limits.}
    \label{fig:Velocities}
\end{figure}

\begin{figure}[t]
    \centering
    \includegraphics[width=0.9\columnwidth]{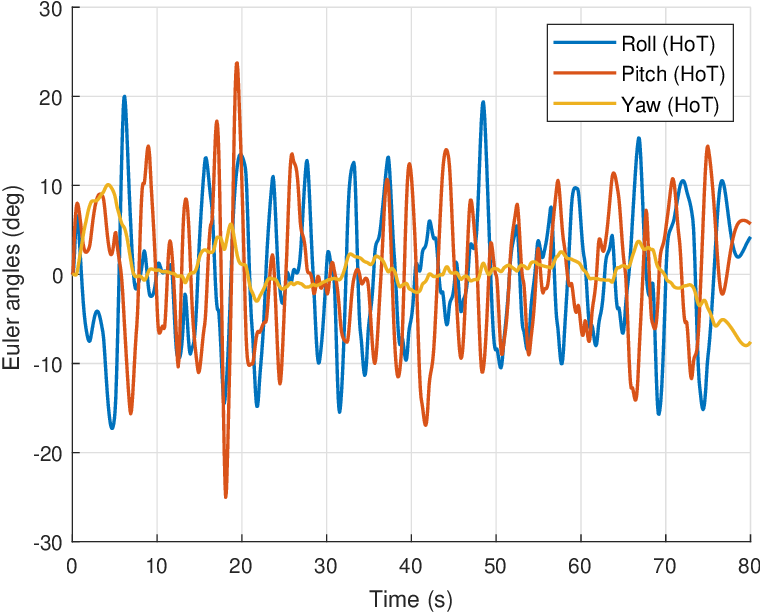}
    \vspace{-0.4cm}
    \caption{Euler angles over time in the HoT scheme, demonstrating smooth attitude behavior.}
    \label{fig:Euler_Angles}
\end{figure}

\begin{figure}[t]
    \centering
     \vspace{0.3cm}
    \includegraphics[width=0.9\columnwidth]{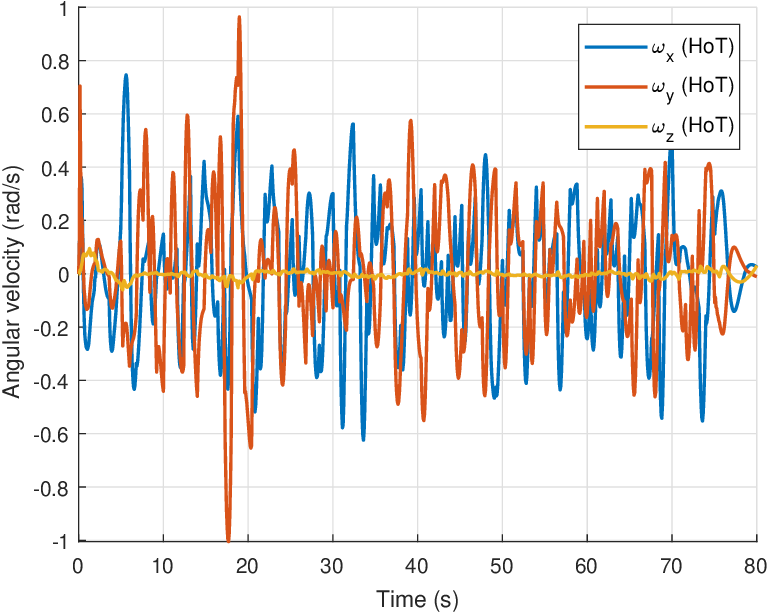}
   
    \caption{Angular velocities over time, showing bounded behavior.}
    \label{fig:Angular_Velocities}
\end{figure}

\begin{figure}[t]
    \centering
    \includegraphics[width=0.9\columnwidth]{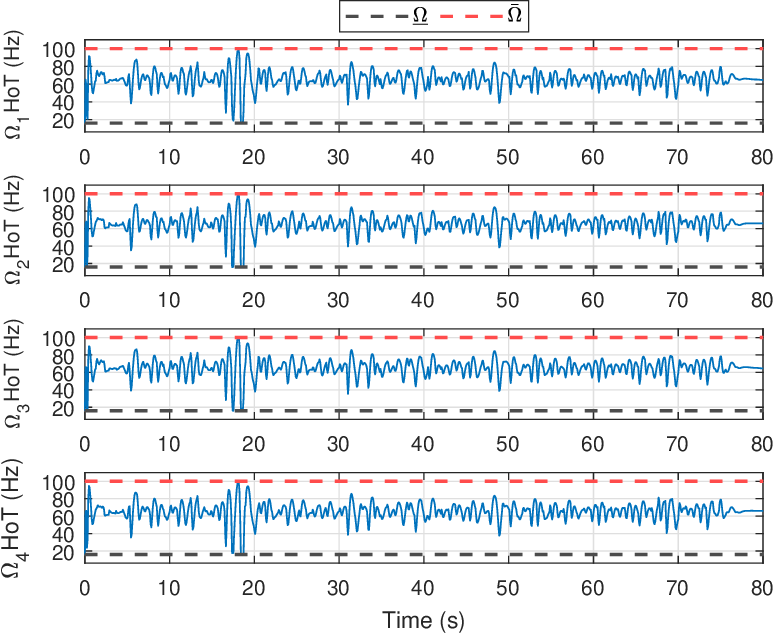}
    \caption{Rotor speeds over time, illustrating smooth actuation within physical limits.}
    \label{fig:Rotor_Speeds}
\end{figure}

\begin{figure}[t]
    \centering
    \includegraphics[width=0.9\columnwidth]{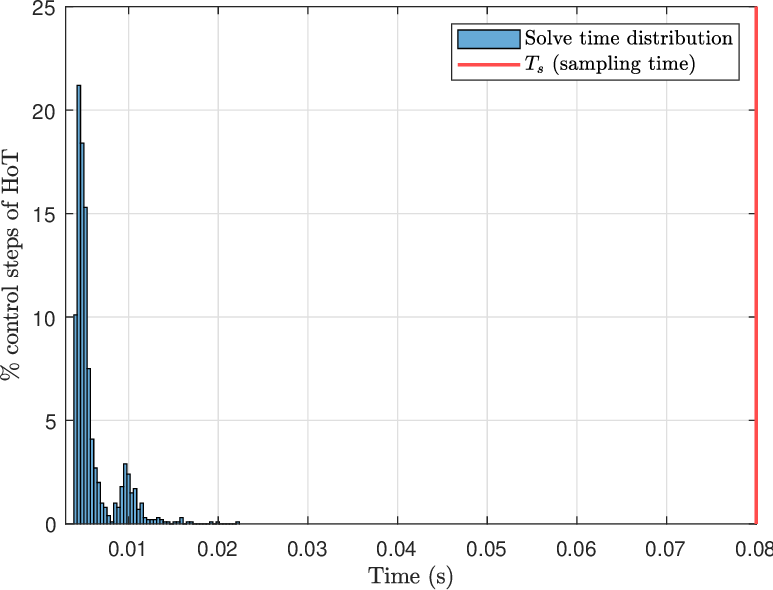}
    \caption{Distribution of the NMPC optimization solve time over all control iterations. The vertical red line denotes the sampling period $T_s$.}
    \label{fig:Histogram_NMPC_Execution}
\end{figure}

This section evaluates the effectiveness of the proposed approach through numerical analysis. We first detail the simulation setup (Section \ref{sec:setup}), followed by an analysis of trajectories (Section \ref{sec:trajectory}), communication performance (Section \ref{sec:comm_perf}), scalability (Section \ref{sec:scalability}), and UAV flight dynamics (Section \ref{sec:dynamics}).

The simulations were performed in MATLAB R2024b using the MATMPC toolbox\footnote{\url{https://github.com/chenyutao36/MATMPC}}~\cite{Chen2019matmpc}, which employs a fixed-step fourth-order Runge--Kutta integrator and qpOASES\footnote{\url{https://github.com/coin-or/qpOASES}} to solve the quadratic programming subproblems arising from the NMPC formulation. All simulations were executed on a desktop computer running Linux Mint $22.3$ (Kernel $6.8.0$), equipped with an Intel\textsuperscript{\textregistered} Core\texttrademark~i$7$-$10700$ CPU ($8$ cores, $16$ threads, up to $4.8$~GHz) and $16$~GB of RAM.



\subsection{Simulation setup and benchmarks} 
\label{sec:setup}

We conducted simulations to assess the performance of our proposed approach by considering an area $\mathcal{A}=[-200;200]^2\times[90;110]$. We consider a group of $I=5$ users deployed with a uniform random distribution for their $x-$ and $y-$coordinates in a square area $(p_{x,i},p_{y,i})\in [100;150]\times [-30;30]$. The BS antenna is positioned at the center of the world frame $O_W$ at an altitude $h^{\mathrm{BS}}=68$ m and is equipped with $N=3$ antenna elements with dimensions $d_{H_0}=d_{V_0}=\lambda/2$, where $\lambda$ is the wavelength of the signal emitted by the BS. The BS, whose antenna gain is $G_A=8$ dB, transmits a signal of wavelength $\lambda=0.15$ m with a transmission power $P_0=23$ dBm and a bandwidth $B_0=5$ MHz. We considered an RIS with $(M_{H}, M_V)=(4,4)$ elements, and the horizontal and vertical separation distances of the RIS elements are $d_{H_R}=\lambda/2$  and $d_{V_R}=\lambda/2$, respectively. We assumed a reference gain at 1 m, $l_0=-30$ dB, and a noise power of $\sigma_0=-100$ dBm. The minimum data rate required to satisfy the users’ QoS is $R_{\min}=1.25$ Mbit/s. The UAV initial position is $\mathbf{p}^{\text{init}}=[-100;-100;h^R]$ and its final destination is $\mathbf{p}^{\text{fin}}=[-100;100;h^R]$ with a desired $z$-coordinate $h^R=100$ m. The hyperparameters for the reference trajectory generation were considered as $L=30000$, $E=8$, and $\gamma=8$. 


The UAV dynamic model parameters and the weights for the NMPC are given in Table~\ref{Tab_NMPC_DM_Params}. The desired orientation, velocity, and angular velocity are $\mathbf{q}_{d,\ell}=[1,0,0,0]^\top$, $\mathbf{v}_{d,\ell}=\mathbf{0}$, and $\boldsymbol{\omega}_{d,\ell}=\mathbf{0}$, respectively, for all time steps $\ell$.  The initial and final states are $\mathbf{x}_I=[\mathbf{p}^{\text{init}},\mathbf{q}^{\text{init}},\mathbf{v}^{\text{init}},\boldsymbol{\omega}^{\text{init}}]^\top$ and $\mathbf{x}_F=[\mathbf{p}^{\text{fin}},\mathbf{q}^{\text{fin}},\mathbf{v}^{\text{fin}},\boldsymbol{\omega}^{\text{fin}}]^\top$, where $\mathbf{q}^{\text{init}}=\mathbf{q}^{\text{fin}}=\mathbf{q}_{d,\ell}$, $\mathbf{v}^{\text{init}}=\mathbf{v}^{\text{fin}}=\mathbf{0}$, and $\boldsymbol{\omega}^{\text{init}}=\boldsymbol{\omega}^{\text{fin}}=\mathbf{0}$. The total flight duration is $T=80$~s.

We evaluate both the dynamic modeling and communication aspects of our different schemes. For the dynamic modeling part, we considered: i) the Horizontal orientation Tracking (HoT) scheme, which corresponds to the proposed tracking scheme in Section \ref{NMPC_Tracking}, for which the tracking parameters are as indicated in Table \ref{Tab_NMPC_DM_Params}; ii) the No orientation Tracking (NoT) scheme, which corresponds to an ARIS with no orientation and angular velocity tracking in the NMPC, meaning that for this scheme $\mathbf{Q}_q=\mathbf{Q}_{q,N_p}=\mathbf{Q}_\omega=\mathbf{Q}_{\omega,N_p}=0$ \eqref{NMPC_Tracking_stage_objective}, and the other tracking parameters are as indicated in Table \ref{Tab_NMPC_DM_Params}; iii) the Straight Line (SL) scheme, in which the ARIS follows a straight-line trajectory from its initial position $\mathbf{p}^{\text{init}}$ to its final position $\mathbf{p}^{\text{fin}}$. The corresponding tracking weights are those reported in Table~\ref{Tab_NMPC_DM_Params}. Regarding the communication aspect, three approaches were considered: i) the proposed BP scheme in Sections \ref{Beamforming_Opt} and \ref{Phase_Shift_Opt}, corresponding to an ARIS where joint beamforming and phase-shift optimization were conducted during the whole mission duration; ii) the P scheme, where only the phase-shift optimization is conducted during the ARIS mission while considering a fixed beamforming vector corresponding to the optimal beamforming at the initial position; iii) the B scheme, with only the beamforming being performed during the flight and a constant phase shift corresponding to the optimal phase shift at the initial position. The SL scheme is considered in conjunction with the BP joint optimization. For the sake of brevity, the resulting SL-BP scheme will hereafter be referred to simply as the SL scheme throughout the remainder of the paper. Regarding the reference trajectory generation, we consider the Proxy Utility Reference Generator (PURG), corresponding to Algorithm~\ref{alg:knn_dijkstra} with the cost defined in~\eqref{Eq:Cost_Computation}, and the Rate Reference Generator (RRG), which corresponds to Algorithm~\ref{alg:knn_dijkstra} using the following cost function: $C_l=R_{\max}-\sum_{i\in\mathcal I}R_i(\mathbf p^l)$, where $R_{\max}$ is the maximum sum-rate among the sampled vertices. The beamforming vector and RIS phase shifts are computed using the closed-form expressions derived in Sections~\ref{Beamforming_Opt} and~\ref{Phase_Shift_Opt}, respectively. Additionally, we consider a Parallel Successive Convex Approximation Reference Generator (PSCA-RG), which employs the PSCA algorithm~\cite{Scutari2014PSCA} to solve a sequence of convex Taylor approximations of the rate function, jointly optimizing the reference trajectory, beamforming vectors, and RIS phase shifts over 40 discrete trajectory points. The resulting trajectory is then densified to match the time discretization adopted by PURG.



\subsection{Trajectory and positional analysis} 
\label{sec:trajectory}

In Fig. \ref{fig:Reference_Trajectory}, the reference trajectory generated by Algorithm \ref{alg:knn_dijkstra} is shown. During the first phase of the mission, the UAV moves towards regions offering a higher data rate, particularly the area close to the users and the BS. In the second phase, the trajectory transitions towards the final destination while still favoring positions with a strong data rate level. Consequently, both objectives of reaching the final position and maintaining high communication quality are achieved, demonstrating the effectiveness of Algorithm~\ref{alg:knn_dijkstra}.

Fig.~\ref{fig:3D_Trajectory} presents the reference trajectory together with the HoT and NoT trajectories in three-dimensional space. While the global motion patterns appear similar, the 3D visualization alone does not clearly quantify the tracking accuracy with respect to the reference. A more precise comparison is provided in Fig.~\ref{fig:3DPosVsTime}, which depicts the time evolution of each position component. From the position-versus-time plots, it can be observed that both schemes start at $\mathbf{p}^{\text{init}}$ and reach $\mathbf{p}^{\text{fin}}$, confirming successful mission completion.  The $x$- and $y$-coordinates exhibit similar behavior in both cases. However, the $z$-coordinate differs more noticeably: the HoT altitude varies between $99.5$\,m and $101.7$\,m, while the NoT altitude varies between $99.9$\,m and $100.1$\,m. The more pronounced variations in the $z$-coordinate observed in HoT, compared to the NoT scheme, are attributed to the orientation and angular velocity tracking imposed on HoT, which is absent in NoT.

Fig.~\ref{fig:Alg1_Vs_PSCA:Ref_Traj} compares the reference trajectories generated by the proposed PURG and the PSCA-RG. Both planners initially steer the UAV toward the communication-favorable region located near the BS and the user cluster before directing it toward the final destination. The PURG trajectory closely follows the overall behavior of the optimization-based PSCA-RG solution despite relying only on the low-complexity proxy cost.

Fig.~\ref{fig:Alg1_Vs_PSCA:Cum_Data_Rate} presents the normalized cumulative sum-rate achieved along the trajectories of Fig.~\ref{fig:Alg1_Vs_PSCA:Ref_Traj}. All curves are normalized by the maximum value attained by the PURG scheme. Both methods exhibit similar growth throughout the mission, confirming that the proxy utility accurately captures the communication objective. Interestingly, PURG achieves a higher cumulative sum-rate than PSCA-RG in this scenario, while requiring only a low-complexity graph-search procedure instead of repeatedly solving convex optimization problems at higher complexity. 

Fig.~\ref{fig:Alg1_PU_Vs_Rate} illustrates the influence of the proxy utility exponent $\gamma$ on the generated trajectory.  For small values of $\gamma$ (e.g., $\gamma=2$), PURG favors shorter trajectories and therefore remains closer to the straight path between the initial and final positions. As $\gamma$ increases, the communication cost becomes more dominant, causing the planned trajectory to move progressively closer to the RRG solution. For $\gamma=8$, the resulting path almost overlaps with the RRG trajectory, indicating that the proxy utility successfully reproduces the behavior of the communication-driven planner.

Fig.~\ref{fig:CumRate_Alg1_PU_Vs_Rate} quantifies the impact of $\gamma$ on the normalized cumulative sum-rate. All curves are normalized using the largest value among the plotted curves. Increasing $\gamma$ consistently improves the cumulative achieved sum-rate because the UAV spends more time in communication-favorable regions. The performance obtained with $\gamma=8$ closely approaches that of the RRG benchmark, while $\gamma=2$ yields the lowest cumulative sum-rate due to its stronger preference for minimizing travel distance. These results demonstrate that $\gamma$ provides an effective mechanism for balancing communication performance and trajectory efficiency.



\subsection{Communication performance comparison} 
\label{sec:comm_perf}

Fig.~\ref{fig:Users_Data_Rate} shows the average users' data rate over time. In the HoT scenario, the average rate evolves smoothly: it increases as the UAV moves toward regions offering stronger channel conditions, reaches a peak around $40$~s, and then gradually decreases as the UAV departs from the high-rate area. The HoT scenario experiences occasional violations of the QoS constraint only at the beginning and the end of the mission, whereas the NoT scenario exhibits frequent QoS violations throughout the mission. The NoT scheme exhibits pronounced fluctuations, reflecting instability induced by uncontrolled orientation changes. The overall parabolic trend is explained by the UAV trajectory: it first approaches a communication-favorable region and later moves away from it during the final phase of the mission. The occasional violation of the minimum rate constraint in the HoT scenario is due to the softened QoS constraint in~\eqref{Const:QoS_OCP}, which allows temporary relaxation to maintain feasibility and ensure mission continuity.
The SL benchmark exhibits a smoother but consistently lower average user rate, remaining close to the minimum QoS threshold because the straight-line trajectory does not exploit communication-favorable regions.

Fig.~\ref{fig:Network_Data_Rate} shows the sum-rate over time. We observe that all NoT-based schemes exhibit fluctuations, while the HoT-based ones demonstrate stability. The SL scheme exhibits a smooth but consistently lower sum-rate than the HoT-BP and NoT-BP schemes, as it spends less time in communication-favorable regions during the mission. From a communication perspective, we note that the BP-based approaches achieve the highest sum rates, followed by the P-based approaches, while the B-based approaches achieve the lowest sum rates. This demonstrates the necessity of joint beamforming and phase-shift optimization over beamforming-only and phase-shift-only optimization. These results also indicate the importance of phase-shift optimization over beamforming optimization, as it plays a key role in signal reflection.

Fig.~\ref{fig:Users_Throughput} illustrates the average cumulative throughput per user over time. The HoT scenario achieves a $13.77\%$ improvement over the NoT scenario, demonstrating the significant impact of orientation tracking on network performance. The shaded regions represent one standard deviation around the average, showing that the users experience similar cumulative throughput due to their clustered spatial distribution. The SL benchmark consistently achieves less throughput than both HoT and NoT, highlighting the performance loss resulting from a communication-unaware straight-line trajectory.

Fig.~\ref{fig:Network_Throughput} shows the cumulative throughput over time. We notice that the HoT schemes, HoT-BP, HoT-P, and HoT-B, outperform their NoT counterparts, namely NoT-BP, NoT-P, and NoT-B, mainly due to the orientation instability in the NoT schemes, which adversely affects both the ARIS motion and the communication channels. The SL, being a BP scheme, consistently achieves lower cumulative throughput than the HoT-BP and NoT-BP schemes, illustrating the performance degradation caused by a communication-unaware straight-line trajectory. Additionally, from a communication perspective, we observe the same trend as in Fig.~\ref{fig:Network_Data_Rate}. The proposed HoT-BP outperforms the HoT-B and HoT-P schemes by at least a factor of $2.5$ and $28\%$, respectively.

Fig.~\ref{fig:Average_Slack} illustrates the evolution of the average slack variable $\mathbf{s}_{\ell}$ throughout the mission. The average is computed over the QoS slack variables of all users at each sampling instant. At the beginning, the slack rapidly increases to compensate for temporary QoS violations caused by the large tilt angles required for the UAV's initial acceleration, which degrade the orientation-dependent RIS channel. During the main flight phase ($t=5$--$75~\mathrm{s}$), the slack is gradually decreased, demonstrating that the penalty term $\|\mathbf{s}_{\ell}\|_{\mathbf{Q}_{sv}}^{2}$ effectively enforces the communication constraints whenever the UAV geometry permits. As the UAV approaches the target, the slack increases again to accommodate the communication degradation associated with the final braking and stabilization maneuvers, thereby preserving optimization feasibility. This behavior highlights the NMPC's ability to balance communication performance and control objectives while maintaining robust operation under challenging flight conditions.



\subsection{Scalability analysis} 
\label{sec:scalability}

In Fig.~\ref{fig:Scale_NRIS}, the throughput of the network is shown as a function of the number of RIS elements by averaging over $5$ network configurations. A minimum rate requirement of $R_{\min}=10$ Kbit/s is used to ensure feasibility, even when the number of RIS elements is small. As can be seen, the HoT-BP scheme outperforms all other schemes for all numbers of RIS elements, demonstrating the effectiveness of the proposed approach. The SL scheme consistently achieves lower throughput than the HoT-BP and NoT-BP schemes across all RIS sizes, while still benefiting from the increased number of RIS elements. Despite employing joint beamforming and phase-shift optimization, the SL scheme consistently underperforms HoT-P. It performs similarly to NoT-P, demonstrating that communication-aware trajectory design is essential for achieving high network throughput. For the maximum number of RIS elements, the performance ranking is HoT-BP, NoT-BP, HoT-P,  SL, NoT-P, HoT-B, and NoT-B. Overall, the network throughput increases with the number of RIS elements for all schemes.

In Fig.~\ref{fig:Scale_NUsers}, the network throughput is shown as a function of the number of users, for all the benchmark schemes, averaged over five network configurations. A minimum user rate of $R_{\min}=1$ Mbit/s is maintained throughout the simulations. Regardless of the number of users, HoT-BP consistently achieves the highest network throughput, followed by NoT-BP, HoT-P, NoT-P, SL-BP, and HoT-B, while HoT-B consistently outperforms NoT-B. The network throughput exhibits only minor variations, as even when the user density increases, the total transmission duration of the TDMA remains unchanged. These results further confirm the importance of jointly optimizing beamforming and RIS phase shifts, as well as explicitly accounting for UAV orientation, to maximize network throughput.



\subsection{Dynamic tracking and actuation performance} 
\label{sec:dynamics}

Fig.~\ref{fig:Tracking_errors} shows the tracking errors of the position, velocity, angular velocity, and the geodesic distance between the UAV orientation and the horizontal orientation over time. While the NoT scheme achieves better position tracking due to fewer constraints, the HoT scheme significantly improves angular velocity and orientation tracking. This behavior is expected, as HoT explicitly enforces attitude regulation, introducing additional coupling in the dynamics that slightly degrades translational accuracy. This trade-off highlights a key limitation of fully coupled control: improving communication-relevant orientation comes at the cost of stricter motion constraints.

Fig.~\ref{fig:Velocities} shows the UAV's velocity profile over time. The velocity evolves smoothly within the imposed bounds, with a gradual deceleration toward zero at the end of the mission. This behavior is consistent with the NMPC design, which anticipates the terminal condition and avoids aggressive braking, indicating good predictive control performance.

In Fig.~\ref{fig:Euler_Angles}, the Euler angles in the HoT scenario remain bounded within a small range and evolve smoothly throughout the flight. This indicates that the controller avoids abrupt attitude changes, which is desirable for both flight stability and communication performance, as large oscillations could degrade the channel.

Fig.~\ref{fig:Angular_Velocities} shows the angular velocity evolution. The angular velocities remain tightly bounded within $[-1,1]$ rad/s, indicating that the controller successfully regulates rotational motion without inducing oscillations or instability.

In Fig.~\ref{fig:Rotor_Speeds}, the rotor speeds remain within actuation limits and vary smoothly over time. The absence of saturation or high-frequency oscillations suggests that the NMPC generates feasible and well-conditioned control inputs, avoiding aggressive commands that could stress the actuators or compromise real-world implementation.

Fig.~\ref{fig:Histogram_NMPC_Execution} shows the histogram of the NMPC computation time over all control iterations. Most optimization problems are solved in less than $20\,\mathrm{ms}$, and all solve times remain well below the sampling period $T_s$ (red line). This demonstrates that the proposed NMPC framework satisfies real-time execution requirements while exhibiting stable and consistent computational performance.

A video of the 3D trajectory of the three schemes can be found in \url{https://youtu.be/TvkwDwx0meg}, and the comprehensive code for this simulation can be found in \url{https://github.com/Akhas2000/ARIS-MATMPC}.



\section{Conclusion}
\label{Conclusion}

In this paper, we studied ARIS deployment for connectivity support in obstructed environments, with a focus on the coupling between UAV motion, RIS orientation, and communication performance. The proposed framework combines orientation-aware channel modeling, multi-rotor UAV dynamics with actuation constraints, closed-form communication updates for beamforming and RIS phase shifts under simplified subproblems, and NMPC-based trajectory tracking. Rather than solving the original joint problem monolithically, the framework addresses it through a decomposition-based strategy that remains computationally tractable and dynamically feasible. Simulation results show that explicit orientation tracking improves throughput and stabilizes the communication performance relative to a free-orientation baseline, while joint beamforming and RIS phase-shift updates outperform partial communication updates. The results also highlight the importance of explicitly controlling the RIS orientation: enforcing orientation tracking leads to a $13.77\%$ improvement in user throughput compared with a free-orientation scenario. Moreover, the joint optimization of beamforming and RIS phase shifts significantly improves network performance, outperforming phase-shift-only and beamforming-only strategies by $28\%$ and up to a factor of $2.5$, respectively. This work demonstrates the strong impact of UAV orientation on ARIS-assisted communications under realistic dynamic constraints. Future work will exploit orientation as an additional degree of freedom for joint communication and trajectory optimization,  and will extend the framework to moving users, a more general multi-user phase-shift design, and energy-aware or security-aware ARIS operation.



\balance
\bibliographystyle{IEEEtran}
\bibliography{Bib_ARIS_V1}

\end{document}